\documentclass[final]{JHEP3}

\usepackage[centertags]{amsmath}
\usepackage{dsfont}
\usepackage{amsfonts}
\usepackage{amsmath}
\usepackage{amssymb}
\usepackage{amscd}
\usepackage{amsthm}
\usepackage{verbatim}

\newcommand{\abs}[1]{\left\vert#1\right\vert}
\newcommand{\nn}{\notag}

\numberwithin{equation}{section}

\def\L{\mathcal{L}}

\def\half{\frac{1}{2}}
\def\s5{\textbf{S}^5}

\def\dds{\partial_1}
\def\ddt{\partial_0}

\def\ddsp{\partial_1'}
\def\ddtp{\partial'_0}
\def\ddz{\partial_{0}}
\def\ddo{\partial_{1}}
\def\dda{\partial_{\alpha}}
\def\ddb{\partial_{\beta}}
\def\ddau{\partial^{\alpha}}
\def\ddbu{\partial^{\beta}}

\def\ddaup{\partial^{\alpha'}}
\def\ddbp{\partial'_{\beta}}
\def\ddbup{\partial^{\beta'}}

\def\denom{\frac{n^2}{r^2}+\frac{4m^2}{l^2}}
\def\denomp{\frac{n'^2}{r^2}+\frac{4m'^2}{l^2}}
\def\summ{\sum_{m=-\infty}^{\infty}}
\def\sumn{\sum_{n=1}^{\infty}}
\def\summN{\sum_{m\neq 0}}

\def\summn{\sum_{m,n}}
\def\summnN{\sum_{(m,n)_{\mathcal{N}}}}
\def\summnD{\sum_{(m,n)_{\mathcal{D}}}}
\def\summnp{\sum_{m',n'}}
\def\summnpD{\sum_{(m',n')_{\mathcal{D}}}}
\def\atE2{\sumn\frac{nq^n}{1-q^n}}
\def\argn{\left(\frac{n\pi \sigma^1}{r}\right)}
\def\argm{\left(\frac{2\pi i m \sigma^0}{l}\right)}

\def\bulkint{\int_{\mathcal{M}}d^2\sigma~}
\def\boundint{\int_{\partial \mathcal{M}}d\sigma^0~}
\def\into{\int d\sigma^1}

\def\argo{\left(\frac{n\pi \sigma^1}{r}\right)}
\def\argop{\left(\frac{n'\pi \sigma^1}{r}\right)}

\def\argop{\left(\frac{n'\pi \sigma^1}{r}\right)}

\def\argc{\left(\frac{n\pi l}{2r}\right)}
\def\eona{\varepsilon^o_{n,1}}
\def\eonb{\varepsilon^o_{n,2}}
\def\eonc{\varepsilon^o_{n,3}}
\def\eond{\varepsilon^o_{n,4}}
\def\eone{\varepsilon^o_{n,5}}

\def\ecna{\varepsilon^c_{n,1}}
\def\ecnb{\varepsilon^c_{n,2}}

\def\ecnd{\varepsilon^c_{n,4}}

\def\ecnf{\varepsilon^c_{n,6}}

\def\ecnbi{\varepsilon^c_{ni_n,2}}

\def\ecndi{\varepsilon^c_{ni_n,4}}

\def\ecnfi{\varepsilon^c_{ni_n,6}}
\def\s{\left(\frac{l}{r}\right)}

\def\fna{f_{n,1}}
\def\fnb{f_{n,2}}
\def\fnc{f_{n,3}}

\def\foa{f_{0,1}}
\def\fob{f_{0,2}}
\def\foc{f_{0,3}}
\def\eonz{\epsilon^o_{0,n}}
\def\ecnz{\epsilon^c_{0,n}}
\def\fnz{F_{n}(l)}
\def\fnzt{\tilde F_{n}(l)}

\title{On the effective theory of long open strings}
\author{Ofer Aharony and Matan Field \\ Department of Particle Physics and Astrophysics, Weizmann Institute of Science, Rehovot 76100, Israel \\  \\
{\tt E-mail:~ Ofer.Aharony@weizmann.ac.il,~Matan.Field@weizmann.ac.il}  }
\abstract{
\noindent We study the general low-energy effective action on long open strings, such as confining strings in pure gauge theories.
Using Lorentz invariance, we find that for a string of length $R$, the leading deviation from the Nambu-Goto energy levels generically occurs at order $1/R^4$
(including a correction to the ground state energy), as opposed to $1/R^5$ for excited closed strings in four dimensions, and $1/R^7$ for closed strings
in three dimensions. This is true both for Dirichlet and for Neumann boundary conditions for the transverse directions, though the worldsheet boundary actions
 are  different. The Dirichlet case is relevant (for instance) for the force between external quarks in a confining gauge theory, and the Neumann case
 for a string stretched between domain walls. In the specific case of confining gauge theories with a weakly curved holographic dual, we compute the coefficient
 of the leading correction when the open string ends on two D-branes, and find a non-vanishing result.
}
\preprint{WIS/10/10-AUG-DPPA}

\begin{document}
\section{Introduction and summary of results} \label{Section1}
Many field theories in $d\geq 3$ spacetime dimensions have stable one dimensional excitations (strings); examples include solitonic strings in theories
like the $d=4$ Abelian Higgs model \cite{Abrikosov:1956sx,Nielsen:1973cs}, and confining strings in pure gauge theories in $d=3,4$.
  Usually those strings have width and are only approximately one (space) dimensional, but in many cases their low-energy effective action can be well described
  by a one dimensional string, with the structural fluctuations considered as extra massive degrees of freedom on the string worldsheet. The state of such a
  string, considered in physical (static) gauge, spontaneously breaks translation invariance in the $(d-2)$ transverse directions, so its embedding orthogonal
 coordinates in spacetime are massless fields on the worldsheet, due to  Goldstone's theorem\footnote{Notice that the string solution actually breaks $3(d-2)$
  symmetry generators, including also the rotations of the parallel directions to the worldsheet with the orthogonal directions. Whereas for internal symmetries
   Goldstone's theorem gives a one-to-one mapping between the number of broken generators and the number of massless Nambu-Goldstone bosons, this is not the case
   for spacetime-dependent symmetries, where a smaller number of Nambu-Goldstone modes is sufficient for the realization of the complete symmetry.}.
   In the absence of any additional symmetries these are generically the only massless modes, and they  interact with the heavy modes. The heavy modes can be
   integrated out to leave an effective action for the massless modes, which is valid up to the energy scale of the (lowest) mass of the integrated heavy modes.
  We assume here that the field theory in the absence of the string has a mass gap so that in the IR limit we are left only with the string's embedding coordinates as massless modes;
  otherwise, the effective action is non-local.
  We also assume that the string is stable, and that at low energies we can ignore interactions between different strings.

      The effective action can be analyzed order by order in the number of derivatives; when the length $R$
       of the string is much longer than the tension length scale, $R>>1/\sqrt{T}$, this derivative expansion is an expansion in $1/R\sqrt{T}$ that is called the `long string'
       expansion. We would like to understand the most general low-energy effective action governing such strings. A reasonable first guess for such an
       effective action is the Nambu-Goto action, whose open string energy levels are known exactly \cite{Arvis:1983fp},
  \begin{align}
  E_n^{o,NG}=TR\sqrt{1+\frac{2\pi}{TR^2}\left[n-\frac{d-2}{24}\right]}~,
  \end{align}
and indeed results from the lattice show a good agreement with this equation (see, for instance,
\cite{Caselle:2002rm}\nocite{Luscher:2002qv,Caselle:2002vq,Caselle:2002ah,
Caselle:2003rq,Caselle:2003db,Caselle:2004jq,Caselle:2004er,Luscher:2004ib,
Caselle:2005xy,Billo:2005iv,Caselle:2005vq,Billo:2005ej,
HariDass:2005we,HariDass:2006pq,HariDass:2007tx,Brandt:2007iw}-\cite{Brandt:2009tc}); a similar agreement is also seen for long closed strings (see, for instance, \cite{Lucini:2002wg}\nocite{Meyer:2004hv,Lottini:2005ya,Caselle:2006dv,Billo:2006zg,Caselle:2007yc,Billo:2007fm,Athenodorou:2007du,
Athenodorou:2007ry,Billo:2007cw,Giudice:2008zk,Giudice:2009di,Athenodorou:2009ms}-\cite{Athenodorou:2010cs}).

 In \cite{Luscher:1980fr, Luscher:1980ac} L\"{u}scher et al. initiated this line of study by writing the leading term in the effective action in static gauge, the free action, and finding the first correction to the classical energy of these strings at order $O(\frac{1}{R})$. Surprisingly (at that time) this was found to be universal, the so called L\"{u}scher term. In \cite{Luscher:2002qv,Luscher:2004ib} L\"{u}scher and Weisz continued this line of study, generalizing it for open and closed strings up to $O(\frac{1}{R^3})$ and including the action on the boundary (with Dirichlet boundary conditions for the transverse coordinates), and found that
 some of the coefficients of the terms in the action are constrained. In \cite{Aharony:2009} further investigation was performed up to $O(\frac{1}{R^5})$,
 for the partition functions on the cylinder and on the torus. All coefficients up to $O(\frac{1}{R^3})$ in the bulk were found to be constrained and equal to
 those of the Nambu-Goto action, as well as all but one of the coefficients at $O(\frac{1}{R^5})$ for $d\geq 4$ \footnote{For the case of $d=3$ the deviating
  term is trivial, and then all the coefficients coincide with those of Nambu-Goto up to (and including) $O(\frac{1}{R^5})$.} . The constraints arise from the
  underlying Lorentz symmetry \cite{Meyer:2006qx,Aharony:2009}. The single unconstrained coefficient was found not to contribute at leading order to the
  partition function of the torus, and it is now believed to be universal as well \cite{Aharony:2010}, although in a less trivial manner. This program explains
  the results from the lattice, where the computed closed string energy levels were found to be very close to those of the Nambu-Goto string. The results above
  also suggest where the first correction to the Nambu-Goto closed string energy levels should be found, but unfortunately this is predicted to occur
  at $O(\frac{1}{R^5})$, a higher order than the one controlled by today's best lattice technology.

 In \cite{Aharony:2009} only the bulk action was discussed. This is sufficient for closed string energy levels, but not for open strings.
 In the present work we generalize the above considerations by considering possible boundary terms up to $O(\frac{1}{R^4})$,
 and we also discuss the case of Neumann boundary conditions for the transverse coordinates\footnote{Dirichlet boundary conditions are relevant, for example,
 for confining strings ending on Wilson loops, while Neumann boundary conditions arise for strings ending on domain walls. In the presence
 of such domain walls, there is also the Nambu-Goldstone mode from translating the domain wall, that we ignore (this field lives in a different
 space, and decouples when the transverse volume of the domain wall is infinite).}. Again,
we find constraints on the coefficients in this action coming from Lorentz symmetry. We find a single non-universal allowed coefficient at this order,
which then gives the leading correction to the Nambu-Goto open string energy levels, at order $1/R^4$. This should be easier to observe on the lattice than
the higher order closed string deviations\footnote{We note that our result is reminiscent of the result of Braaten et al. \cite{Braaten:1986bz,Braaten:1987gq}, who
compute the static potential for the string action with the rigidity term \cite{Polyakov:1986cs}, in addition to the Nambu-Goto term, in the large $d$ limit,
and find the first correction to Nambu-Goto to be at the $1/R^4$ order, after expanding their results in $1/R$.
However, the two results are probably unrelated; the results of \cite{Braaten:1986bz,Braaten:1987gq} are obtained in the large $d$ limit, and only then
expanded in $1/R$, but the large $d$ and large $R$ limits generally do not commute, although they happen to commute in the Nambu-Goto case,
since the radius of convergence for the $1/R$ expansion goes to infinity when $d$ goes to infinity. This is also clear from the fact that the result
of \cite{Braaten:1986bz,Braaten:1987gq} is non-analytic in the string tension, as well as in the rigidity coefficient, so it is probably non-perturbative in $1/R$.
We stress that our work is inherently perturbative in $1/R$, and thus does not imply anything outside the radius of convergence of the $1/R$ expansion. We also note that within the $1/R$ expansion, and at the order we work at, the rigidity term is trivial (can be removed by field redefinitions).}.
Our main result is that for Dirichlet
boundary conditions on the transverse directions
the only allowed non-constant boundary term up to four-derivative
order is of the form $b_2 \ddt \dds X\cdot \ddt \dds X$, with an arbitrary coefficient $b_2$. This leads to a correction to
the Nambu-Goto result for the open string ground state energy of the form $\delta E_0 = -b_2 \pi^3 (d-2) / 60 R^4$ (measured in units of the string tension). For
Neumann boundary conditions the allowed boundary terms up to
four-derivative order are $\mu[1 + \frac{1}{2} \ddt X \cdot \ddt X - \frac{1}{8}
(\ddt X \cdot \ddt X)^2] + a_2 \ddt^2 X \cdot \ddt^2 X$,
with arbitrary coefficients $\mu$ and $a_2$.

As in \cite{Aharony:2009}, we test our form of the effective action by a holographic computation in a confining
gauge theory that has a dual string theory description  (by the AdS/CFT correspondence \cite{Maldacena:1997re}) as a superstring on a weakly curved background.
 The computation is done for a long string stretched between two D-branes (sitting in the confining region), and as
 expected we obtain non-zero values for the allowed non-universal coefficients; we treat both the Dirichlet and Neumann cases.
 Note that while the computation with Neumann boundary conditions for the transverse fluctuations refers to a long string stretched between domain walls, our computation with Dirichlet boundary conditions for the transverse fluctuations is not exactly the same as
 a correlator of Wilson loops, which is what is usually measured on the lattice in this context.
 We leave for future work the holographic computation of the latter, but we expect it to
 give similar results.

 In section \ref{Section2} we write the most general effective action for Dirichlet and Neumann
  boundary conditions up to $O(1/R^4)$, in static gauge and in the long string effective action.
  In section \ref{Section3} we compute the corrections to the cylinder partition function coming from this general action
  in both cases. In section \ref{Section4} we derive the most general allowed form for the cylinder partition function,
  by writing it as a sum over propagating states in the open and closed channels, and compare it with our previous results.
  This leads to constraints on the coefficients in the effective action, and also to expressions for the corrections to the energy levels.
  The energy levels of the effective string action can also be computed directly in a Hamiltonian formalism \cite{Aharony:2010a}, which gives the same results.
  In particular, we find the lowest order correction term to the Nambu-Goto string, which arises at order $1/R^4$ and with one arbitrary free coefficient, both in the Dirichlet and Neumann cases.
  In section \ref{Section5}, which is independent of sections \ref{Section3} and \ref{Section4}, we derive the same constraints in a much simpler and direct way, by demanding invariance of the action under the implicit parts of the Lorentz group.
  Finally, in section \ref{Section6} we consider holographic models of confining gauge theories with both boundary conditions, and by integrating out the massive modes we compute the effective action
  for the confining string. We verify the constraints, and compute the allowed free parameters in each case, which are set in this framework by the holographic geometry. In the appendices we present all our notations, technical details and computations.

\section{The effective action}\label{Section2}

The effective action we consider is a derivative expansion around a long string solution. To compute the open string partition function we take a
 cylindrical worldsheet, wrapping a periodic compact dimension $X^0$ of length $L$ and stretching with length $R$ along one flat non-compact direction $X^1$,
 and we fix the worldsheet diffeomorphism gauge freedom in static gauge by choosing $\sigma^0=X^0~,~\sigma^1=X^1$.
 The operators in the action then include only the massless modes $X^i$ ($i=2,...,d-1$, where $d$ is the number of spacetime dimensions),
 which are the transverse fluctuations of the string around the classical solution; these fields must always show up with derivatives to keep
  the translation invariance.
 The effective action in this gauge manifestly preserves
  the transverse rotation symmetry $SO(d-2)$, as well as the Lorentz symmetry on the worldsheet $SO(1,1)$ (broken by the boundary) and so we build
  operators that are invariant under these symmetries. In the convention that the $X$'s are dimensionless and in any computation,
  the contributions from higher dimensional operators in the action are suppressed by inverse powers of the long string length (L or R).
  We will thus look for all possible independent terms in the action, order by order in the number of derivatives.
  Since worldsheet coordinates have dimensions of length, a bulk term has the same order as a boundary term with one less derivative.

   It is well known that terms in the action that are proportional to the equation of motion (e.o.m.) or its derivatives do not contribute in perturbation theory
   and can be swallowed by field redefinitions. When working at a fixed order in the derivative expansion this allows us to use in the action the e.o.m. of the free theory, since corrections to this will generate terms at higher orders, which we can ignore since we classify the most general terms at each order anyway. Alternatively, we prove directly in perturbation theory in appendix \ref{appendixD} (and specifically on the cylinder, with both boundary conditions) that terms in the action that are proportional to the free e.o.m., or its derivatives, give no contribution to the partition function in perturbation theory, and thus we ignore terms of this kind. We work all along in Euclidean signature.

 The simple constant term in the bulk,
\begin{align}
T\bulkint = TLR~,
\end{align}
 gives us the definition of the string tension $T$. From here on we rescale our worldsheet coordinates with the square root of the tension, such that they
 are dimensionless coordinates ranging in $\sigma^0\in[0,l]~,~\sigma^1\in[0,r]$, with $l\equiv L\sqrt{T}~,~r\equiv R\sqrt{T}~>>~1$ the parameters for the
 long string expansion. Any additional term contributes with appropriate powers of the long string lengths ($l,r$) from the worldsheet derivatives, and we
 denote by `order k' general terms in the bulk of the form $~\bulkint \partial^{k+2} X^{2n}~$, and corresponding terms on the boundary
 $~\boundint \partial^{k+1} X^{2n}~,$ and by $S_k$ ($S'_k$) the bulk (boundary) action at order $k$.
  \textbf{Up to order 0} we can only write the free action,
 \begin{align}\label{eqDA0}
  &S_0= l(r+2\mu)+ \half\bulkint \partial_{\alpha} X \cdot \partial^{\alpha} X~,
  \end{align}
 where we scale our fields to have a canonical kinetic term. Since we have set the fields to be dimensionless as well, this will make all couplings that
 we write for higher level terms dimensionless, and so everything in our action is expressed in units of appropriate powers of the string tension.
 The constant term on the boundary gives the end-points of the
string a static ``mass'' \footnote{This is not really a mass in space-time, since the
end-point is always fixed in the $X^1$ direction. In the case of Neumann boundary conditions
for some of the transverse directions, the end-points are free to move in these directions,
and then $\mu$ behaves as an end-point mass for the motion in these directions. Note that the
effective string action is not valid for fully dynamical end-points, since in that case nothing
prevents the string from becoming short.}; note that the boundary consists of two
 components \footnote{More generally one can assign two independent couplings for the boundary action on the two disconnected boundaries
  \begin{align}
  b^iS'^i+b^fS'^f=b^i\int d\sigma^0\mathcal{O}|_{\sigma^1=0}+b^f\int d\sigma^0\mathcal{O}|_{\sigma^1=R}~,
   \end{align}
   since the string can end on different objects at its two ends. This can also be rewritten using
   \begin{align}
   b^{\pm}S'^\pm= b^{\pm}\left(\int d\sigma^0\mathcal{O}|_{\sigma^1=0} \pm \int d\sigma^0\mathcal{O}|_{\sigma^1=R}\right)~.
    \end{align}
      We work at first order in the higher derivative operators, such that $S'^-$ does not contribute, and so we will only consider $S'=S'^+$. Of course, in the case of different boundary conditions at the two ends this statement is irrelevant since we could have different operators on different boundaries; we do not analyze this case here, although it can be done in a similar manner.},
  so that the mass for each end-point (in units of the square root of the string tension) is $\mu$.

We now write down the general terms in the effective action, allowed by the manifest symmetries, order by order, in the cases of Dirichlet and Neumann boundary conditions. In the next sections
we will see how some of these terms are constrained by Lorentz invariance.

  \subsection{Dirichlet boundary conditions}

  In the case of Dirichlet boundary conditions, which we take without loss of generality to be $X^i=0$ at both ends, pure $\sigma^0$-derivatives vanish
  on the boundary ($\partial_0^n X=0$).

  \textbf{ At order 1} the general allowed form is schematically $\boundint \partial^{2} X^{2}$ and the only possible term in the Lagrangian is
  therefore \cite{Luscher:2002qv}\footnote{Note the difference in the definition of $b_1$, as well as $c_2,c_3$ in \eqref{eqDA2}, by a factor of $4$ from that of
  \cite{Luscher:2002qv,Luscher:2004ib}.}
\begin{align}\label{eqDA1}
 \L'_1 = b_1\dds X \cdot \dds X.
\end{align}

\textbf{At order 2}, terms in the bulk are of the form $\bulkint \partial^{4} X^{k}~$ for  $k=2,4$; this action was written already in
\cite{Luscher:2004ib,Aharony:2009}. For $k=2$ all terms include the equation of motion or they are related to such terms through integration by
parts\footnote{It is true that here integration by parts does not give equivalent terms immediately because of the boundary, but in any case it does not
give new terms that are not already taken into account when writing the boundary terms by themselves.}.
There are two $k=4$ terms,
\begin{align}\label{eqDA2}
 \L_2
 = c_2 (\dda X \cdot \ddau X)(\ddb X \cdot \ddbu X) + c_3 (\dda X \cdot \ddb X)(\ddau X \cdot \ddbu X)~.
\end{align}
On the boundary the general term is of the form $\boundint \partial^{3} X^2~$  and there are two such possible terms,
 \begin{align}\label{naive L'2}
  \ddt\dds X \cdot \dds X~,~\dds^2 X \cdot \dds X~,
\end{align}
both of which are trivial. In the second term one can transform the $\dds^2$ into $\ddt^2$ by the e.o.m. and then it is trivial by the boundary condition.
The general rule is that \textbf{an operator on the boundary is non-vanishing only if each $X$ has an odd number of $\dds$'s.}\footnote{The above statement,
which is the generalization of the e.o.m. on the boundary, $\dds^2X^i=0$, is also transparent from the form of the propagator,
which obeys $\dds^{2n} G_{\mathcal{D}}|_{\partial \mathcal{M}}=0$ (see \eqref{calDprop} for details).}
The first term in \eqref{naive L'2} is a total time derivative and thus it is trivial. Note that in general, terms with an odd number of $\ddt$'s do not
contribute by their one-vertex function, but can contribute through higher vertex functions.

Using the rule above we easily classify all possible boundary terms at all orders. Since $\dds^2=-\ddt^2$ by the free e.o.m, and since any $X$ should come with an odd number of $\dds$'s, each $X$ should have a single $\dds$ and an arbitrary number of $\ddt$'s, determined by the term's order; at order $k$ the
boundary term should have $k+1$ derivatives. By demanding translation and rotation invariance\footnote{In $d=3$ we should demand spacetime parity for this claim.}
for the transverse coordinates one should have an arbitrary even number of $X$'s that is smaller or equal to the number of derivatives. We note that some
different configurations are dependent by integration by parts and one should carefully choose the independent terms. Of course we should also exclude terms
that are total derivatives.

\textbf{At order 3} we find then
 \begin{align}\label{eqDA3}
  \L'_3= b_2(\ddt\dds X\cdot \ddt\dds X)+b_3(\dds X\cdot \dds X)^2,
\end{align}
and we will see that this gives the leading correction to Nambu-Goto, so we will not discuss higher orders here;
note that \textbf{at order 4} there are possible bulk terms \cite{Aharony:2009}, but no boundary action.
 Up to order $3$ the general effective Lagrangian on the boundary is thus
\begin{align}
\mathcal{L'}=&[\mathcal{L'}_{-1}]+[\mathcal{L'}_1]+[\mathcal{L'}_3] \notag \\
=&[\mu]+[b_1\dds X\cdot \dds X]+[b_2 \ddt\dds X\cdot \ddt\dds X+b_3(\dds X\cdot \dds X)^2]~.
\end{align}

\subsection{Neumann boundary conditions}

Here a similar all-order classification can be made. The boundary condition $\dds X^i|_b=0$  translates into the rule that
\textbf{an operator on the boundary is non-vanishing only if each $X$ has an even number of $\dds$'s.}
The use of the equation of motion together with the above rule allows us to disregard all $\dds$'s, and have only $\ddt$'s.
The general action on the boundary up to order 3 is then
\begin{align}\label{eqNA}
\mathcal{L}'=&[\mathcal{L}'_{-1}]+[\mathcal{L}'_1]+[\mathcal{L}'_3] \notag \\
=&[\mu]+[a_1\ddt X\cdot \ddt X]+[a_2 \ddt^2 X\cdot \ddt^2 X+a_3(\ddt X\cdot \ddt X)^2]~,
\end{align}
and the bulk action is as above. The first term gives a mass for the end-points, and the second term is their kinetic term. Note that in this case we could translate the boundary terms into
modified boundary conditions, but we use here a formalism in which the boundary conditions
are fixed.

\section{The partition function}\label{Section3}

We call `level $k$' a contribution to the partition function $Z^{(k)}$ that corrects the free partition function at order $O(r^{-j}l^{-k+j})$,
for some integer $j$, so that $Z=Z^{(0)}+Z^{(1)}+...~$, and $Z^{(k)}/Z^{(0)}\sim O(r^{-j}l^{-k+j})$. In these notations an order $k$ term in the action
 first contributes to $Z$ through its one-vertex function at level $k$. We now write down these contributions level by level. The notations and functions that we use
in this section are similar to those of \cite{Aharony:2009} and are summarized in appendix \ref{appendixA}. We state in this section all the results (old and new),
and in appendix \ref{appendixE} we present their detailed derivations.

\subsection{Dirichlet boundary conditions}

We begin with the case of Dirichlet boundary conditions for the transverse coordinates.
 The partition function of the free action \eqref{eqDA0} is known for many years to be \cite{Dietz:1982uc}
\begin{align}\label{eqDPF0}
 Z_{\mathcal{D}}^{(0)}=\int \mathcal{D}X \exp[-S_0]= e^{-l(r+2\mu)}\eta(q)^{2-d}~,
\end{align}
with the definitions:
\begin{align}\label{}
 q\equiv e^{2\pi i \tau}=e^{-\frac{\pi l}{r}} ~~,~~\tau\equiv i\frac{l}{2r} ~.
\end{align}
This computation (see \eqref{calDPF0} below) uses the $\zeta$-function regularization \cite{Dietz:1982uc} to
regulate\footnote{This technique actually performs at once both regularization and renormalization, i.e. it turns an infinite sum to a finite result.
The infinities come from the infinite energy of the worldsheet vacuum, and its renormalization amounts to the renormalization of the string
tension \cite{Nesterenko:1997ku}. It was shown in \cite{Caselle:1996kd} that the finite result is universal and independent of the various possible regularization schemes.}
 the infinite product obtained from the determinant of the Laplacian on the cylinder.

\textbf{At level 1} the additional contribution to the partition function coming from the action \eqref{eqDA1} is \cite{Luscher:2002qv} at leading order
\begin{align}\label{eqDPF1}
 &Z=\int DX e^{-S_0-S'_1}\simeq Z^{(0)} (1- \langle S'_1 \rangle )=Z^{(0)}+Z^{(1)}~, \notag \\
 &\langle S'_1 \rangle =b_1 (d-2)\boundint \dds \ddsp G_{\mathcal{D}} =-b_1 (d-2)\frac{\pi l}{6 r^2} E_2(q)~,
\end{align}
where $G_{\mathcal{D}}(\sigma,\sigma')$ is the massless propagator on the cylinder with Dirichlet boundary conditions \eqref{calDprop};
for the details see \eqref{calDPF1} below. As shown in \cite{Luscher:2004ib} and reviewed below, when expanding this result in the closed channel
and comparing it with the general form of the partition function, the only consistent value for the coupling is $b_1=0$. We will therefore ignore
this term from here on, also in higher level computations, where it could have contributed through a multi-vertex correlation function.

\textbf{At level 2} we find a contribution to the partition function coming from \eqref{eqDA2}, given by $Z^{(2)}=-Z^{(0)} \langle S_2 \rangle$
(we ignore the $\langle S_1'^2 \rangle$ contribution), and it was calculated (see \eqref{calDPF2} below for details) in \cite{Luscher:2004ib,Aharony:2009} (in three dimensions it was already calculated in \cite{Dietz:1982uc,Caselle:2002rm}),
\begin{align}\label{eqDPF2}
&\langle S_2 \rangle = (d-2)\{[(d-2)c_2+c_3]I_1+[2c_2+(d-1)c_3]I_2\}~,
\end{align}
with:
\begin{align}\label{}
 &I_1=\bulkint \dda \ddaup  G_{\mathcal{D}}~ \ddb \ddbup G_{\mathcal{D}}=\frac{2\pi^2 l}{r^3}H_{2,2}(q)~, \notag \\
 &I_2=\bulkint \dda\ddbp G_{\mathcal{D}} ~\ddau \ddbup G_{\mathcal{D}}=\frac{\pi^2 l}{288r^3}E_4(q)~.
\end{align}

 \textbf{ At level 3} the only contribution to the partition function comes from \eqref{eqDA3} and is $ Z^{(3)}= - Z^{(0)}\langle S'_3 \rangle$
 (ignoring the $\langle S'_1 S_2 \rangle$ contribution). This is given by (see \eqref{calDPF3} for details)
\begin{align}\label{eqDPF3}
&\langle S'_3 \rangle=  (d-2)[b_2I_3+db_3I_4]~,
\end{align}
with:
\begin{align}\label{s3int}
 & I_3=\boundint \ddt \dds \ddtp \ddsp G_{\mathcal{D}} ~=~-\frac{\pi^3l}{60r^4} E_4(q)~, \notag \\
 & I_4=\boundint (\dds \ddsp G_{\mathcal{D}})^2~=~\frac{\pi^2l}{72r^4} E_2(q)^2~.
\end{align}

\subsection{Neumann boundary conditions}

With Neumann boundary conditions for the transverse directions the propagator is \eqref{calNprop}, as derived in the appendix.
In addition to switching sines with cosines (relative to the Dirichlet case), since the transverse string position is no longer fixed at its ends in this case, there is an extra piece which is obtained by considering also the spatially constant $n=0$ terms (but disregarding the zero mode $n=m=0$).

At \textbf{level 0} we compute the partition function (see \eqref{calNPF0}) to be
\begin{align}\label{eqNPF0}
&Z_{\mathcal{N}}^{(0)}=\mathcal{V}_{\perp}\left(\frac{r}{2\pi l}\right)^{\frac{d-2}{2}}e^{-l(r+2\mu)}\eta(q)^{2-d}~.
\end{align}
The difference from the Dirichlet case is due to the spatially constant modes and in particular the zero modes which were absent before.
$\mathcal{V}_{\perp}$ is the volume of the transverse $X$-space (which is dimensionless and related to the volume $V_\perp$ of the real positional space by $\mathcal{V}_{\perp}=V_{\perp}T^{\frac{d-2}{2}}$).

\textbf{At level 1} we have a contribution to the partition function coming from \eqref{eqNA}, given by $Z^{(1)}=-Z^{(0)}\langle S'_1 \rangle$
with (see \eqref{calNPF1} for details)
\begin{align}\label{eqNPF1}
 &\langle S'_1 \rangle  =  a_1 (d-2) \boundint \ddt \ddtp G_{\mathcal{N}}=
 -2a_1 (d-2)\left\{1-\frac{\pi l}{12r}E_2(q)\right\}\frac{1}{r}~.
\end{align}
We will later show that in this case $a_1$ is proportional to $\mu$, the mass parameter of order $(-1)$.
When $\mu \neq 0$, the higher level contributions to the partition function are then more complicated than before,
\begin{align}
Z&=Z^{(0)}\left[1-\langle S'_1+S_2+... \rangle+\half\langle (S'_1+S_2+...)^2 \rangle+... \right] \\
&=Z^{(0)}\left[1-\langle S'_1\rangle-\left(\langle S_2\rangle-\half \langle S_1'^2\rangle \right)-\left(\langle S'_3\rangle-\langle S'_1 S_2\rangle+\frac{1}{6}\langle S_1'^3\rangle\right)-...\right],\nonumber~
\end{align}
and they involve multi-vertex correlation functions which are harder to compute. For our purposes the terms that are
independent of $\mu$ and are easy to compute are sufficient, and so at higher levels these are the only ones we will compute.

 \textbf{At level $2$} the relevant contribution to the partition function coming from \eqref{eqDA2} is then $Z^{(2)}\approx-Z^{(0)}\langle S_2\rangle$,
 where the similarity symbol denotes terms that are independent of $\mu$. The structure of contractions and integrals is unchanged from
 the Dirichlet case \eqref{eqDPF2},
 \begin{align}
&\langle S_2 \rangle = (d-2)\{[(d-2)c_2+c_3]I_1+[2c_2+(d-1)c_3]I_2\}~,
\end{align}
but their values do change due to the change in the propagator (see \eqref{calNPF2} for the detailed calculation):
\begin{align}\label{eqNPF2}
 &I_1=\bulkint \dda \ddaup  G_{\mathcal{N}}~ \ddb \ddbup G_{\mathcal{N}}=\frac{2\pi^2 l}{r^3}H_{2,2}(q)+\frac{1}{rl}~,  \\
 &I_2=\bulkint \dda\ddbp G_{\mathcal{N}} ~\ddau \ddbup G_{\mathcal{N}}=\frac{\pi^2 l}{288r^3}E_4(q)-
 \frac{\pi}{12r^2}E_2(q)+\frac{1}{rl}~. \nn
\end{align}

 \textbf{At level $3$} the relevant (independent of $\mu$ and $a_1$) contribution to the partition function coming from \eqref{eqNA}
is $Z^{(3)}\approx-Z^{(0)}\langle S'_3 \rangle$ with (see \eqref{calNPF3} for details):
\begin{align}\label{eqNPF3}
&\langle S'_3 \rangle~=~(d-2)[a_2 I_3+a_3 d I_4]~,
\end{align}
\begin{align}
 & I_3=\boundint \ddt^2 \partial_0^{'2} G_{\mathcal{N}} ~=~\frac{\pi^3l}{60r^4} E_4(q)~,\nn\\
  & I_4=\boundint (\ddt \ddtp G_{\mathcal{N}})^2~=~\frac{\pi^2l}{72 r^4} E_2(q)^2-\frac{\pi}{3r^3}E_2(q)+\frac{2}{r^2l}~.
 \end{align}
\section{Comparison with a general ansatz}\label{Section4}

In the limit of a long and
stable (non-interacting) string, the partition function of this string over some surface must have an interpretation in terms of propagation
of physical string states along this surface, and its action needs to be diffeomorphism invariant.
The cylindrical worldsheet has two different interpretations depending on the choice of time direction. If $X^0$ is chosen for time then the interpretation is that of an open string winding time periodically, and the partition function is a thermal one for the open string; this is called the `open channel'.
 When $X^1$ is chosen for time, the partition function is just a propagation amplitude for a closed string between two boundary states,
 and this is denoted the `closed channel'.
 As mentioned above, fixing the diffeomorphism invariance
by the static gauge leaves manifest only an $SO(1,1)\times SO(d-2)$ part of the complete Lorentz group, and we have built the most general effective action
symmetric under this part. Demanding then the complete symmetry (i.e. including the rotation of parallel directions to the worldsheet with
transverse directions)\footnote{Strictly speaking this part of the symmetry, as well as the Lorentz symmetry on the worldsheet, are only preserved
when the worldsheet is on the plane. However, since the action is local, the symmetry is also manifest on the cylinder and the torus, when not broken
by boundary conditions; in the latter case only the subgroup of Lorentz that is preserved by the boundary conditions can be used as a symmetry.}
will constrain the coefficients in this action, and there are several ways to implement this demand and then to obtain the constraints.
 In this section we use the old method \cite{Luscher:2004ib,Aharony:2009},
  where the complete Lorentz symmetry is used to construct a general ansatz for string partition functions. The general ansatz \textbf{in one of these channels} explicitly uses the non-manifest part of Lorentz, as shown below,
 and thus when compared to the computed partition function, it will lead to the constraints on the coefficients in the string action. We will see that for each boundary condition a different channel will encode this implicit part of the Lorentz symmetry. The comparison of the computed partition function with the general ansatz in the open (closed) channel will also give the corrections to the open (closed) string energies. A different method to find the
 Lorentz invariance constraints \cite{Aharony:2010} is explained in the next section.

\subsection{The general form of the partition function}

 We begin with \textbf{Dirichlet} boundary conditions, $\ddt X^i|_b=0$, and with the \textbf{open channel} ($X^0$ is time).
 The open string that runs in the loop is attached at its ends and there is no transverse momentum; namely, the zero modes of the $X$'s are fixed.
 Thus, if we expand the string in modes and denote by $|n;X^0\rangle$ the quantum state of a string of length $R$ in the $n$'th oscillatory
 state\footnote{In a general theory of strings these are no longer given by the usual free field Fourier modes, but working in the canonical formalism
 one can always diagonalize the Hamiltonian on the worldsheet to obtain its eigenstates, which include the oscillatory energy part and the kinetic energy
 for the center of mass coordinates. We assume here that those two parts can be consistently separated, meaning that the Hilbert space is block diagonal,
 and when having also transverse momentum $| n,P_T\rangle \equiv | n \rangle \otimes | P_T \rangle$.
 Note that we consider the worldsheet energies although we are actually interested in the energy levels of the string in spacetime.
 However, in the static gauge the two coincide.} at time $X^0$, the thermal partition function sums in the loop only over all oscillatory states,
\begin{align}
&Z_{\mathcal{D}}^{[o]}(L,R)=\sum_{n}\langle n;L|n;0\rangle=\sum_{n}\langle n|e^{-\mathcal{H}^oL}| n\rangle =\sum_n e^{-E^o_n(R)L}~,
\end{align}
where $\mathcal{H}^o$ is the worldsheet Hamiltonian of the open string.
 For more general boundary conditions the open string can have also a (transverse) kinetic energy for its center of mass so that its total energy is really
 $E^o_{n,P_{T}}(R)$, and here we mean $E^o_n(R)\equiv E^o_{n,P_{T}=0}(R)$, the energy of the open string with no transverse momentum.

\textbf{In the closed channel} ($X^1$ is time) the \textbf{Dirichlet} boundary condition states that at the initial and final times the whole string is located
at a single point in the transverse space; in particular its center of mass is located at that point\footnote{Notice that the boundary state overlaps with
generic states allowed by the symmetries.}. Now the string quantum state also includes the state for the position (or momentum) of its center of mass
in the transverse space (or momentum space), and this will be denoted by $|n,X_{\perp};X^1\rangle$ or $|n,P_{T};X^1\rangle$, respectively. Denoting the boundary state
by $|B\rangle$ (which is the state of fixed $X_{\perp}$) and choosing no transverse separation between the two edges, we find
\begin{align}\label{closed channel derivation}
Z_{\mathcal{D}}^{[c]}(L,R)&=\langle B,X_{\perp};R|B,X_{\perp};0\rangle=\langle B,X_{\perp}|e^{-\mathcal{H}^cR}|B,X_{\perp}\rangle \notag \\
&=\sum_n \int \frac{d^{d-2}P_{T}}{(2\pi)^{d-2}}\frac{E_{n}^c(L)}{E_{n,P_{T}}^c(L)} \langle B,X_{\perp}| n,P_{T}\rangle \langle n,P_{T}|B,X_{\perp}\rangle
e^{-E_{n,P_{T}}^c(L)R}\notag \\
&=\sum_n \int \frac{d^{d-2}P_{T}}{(2\pi)^{d-2}}\frac{E_{n}^c(L)}{E_{n,P_{T}}^c(L)} \langle B| n\rangle \langle n|B\rangle e^{iP_{T}\cdot (X_{\perp}-X_{\perp}) }e^{-E_{n,P_{T}}^c(L)R}\notag \\
&=\sum_n \abs{v_n(L)}^2 \int \frac{d^{d-2}P_{T}}{(2\pi)^{d-2}} \frac{E_{n}^c(L)}{E_{n,P_{T}}^c(L)} e^{-E_{n,P_{T}}^c(L)R}~.
\end{align}
 In the first equality we have put in a complete set of states in a Lorentz invariant manner
 \cite{Meyer:2006qx}\footnote{This follows the usual considerations (see for example \cite{Peskin:1995ev}, page 23), except that now the particle is
 replaced with a string which has also the oscillatory energy and we find the correct normalization through comparing with the original computation method
 of \cite{Luscher:2004ib}. The Lorentz invariant measure is then
\begin{align}\label{LIMeasure}
\int \frac{d^{d-2}P_{T}}{(2\pi)^{d-2}}\frac{E_n(L)}{E_{n,P_{T}}(L)}~,
\end{align}
 and correspondingly
 \begin{align}\label{DeltaOperator}
 \langle n,P_{T}|m,P'_{T}\rangle=\delta_{nm}\frac{E_{n,P_{T}}(L)}{E_n(L)}(2\pi)^{d-2}\delta^{(d-2)}(P_{T}-P'_{T})~,
 \end{align}
  which we use below.}, and we denote by $E_{n,P_{T}}^c$ the closed string energies with
  transverse momentum $P_T$, and by $v_n(L)\equiv \langle n|B\rangle$ the overlap of the
  string energy eigenstates with the boundary state.
Using the relativistic dispersion relation \cite{Meyer:2006qx}
\begin{align}\label{Dispersion relation}
E_{n,P_{T}}^c(L)=\sqrt{E^c_n(L)^2+P_{T}^2}~,
\end{align}
as well as the integral formula
\begin{align}\label{Integral formula1}
 &\int_0^{\infty} dx~\frac{x^n}{\sqrt{x^2+\alpha^2}}e^{-\sqrt{x^2+\alpha^2}}=\frac{1}{\sqrt{\pi}}\Gamma\left(\frac{n+1}{2}\right)(2\alpha)^{\frac{n}{2}}
 K_{\frac{n}{2}}(\alpha)~,
\end{align}
we solve explicitly the integral over the transverse momentum in (\ref{closed channel derivation}) to find
\begin{align}\label{closed channel expansion}
Z_{\mathcal{D}}^{[c]}(L,R)=2R^{2-d}\sum_n \abs{v_n(L)}^2 \left(\frac{E^c_n(L)R}{2\pi}\right)^{\frac{d-1}{2}}K_{\frac{d-3}{2}}(E^c_n(L)R)~,
\end{align}
where $K_{\nu}(x)$ are the modified Bessel functions of the second type. Note that in (\ref{Dispersion relation}) we use the implicit part of the
(explicitly) unbroken Lorentz group that rotates the $X^1$ direction with the transverse directions; this
part is not broken since $X^1$ also has Dirichlet boundary conditions. The same result was derived in \cite{Luscher:2004ib} by a slightly different method.

The \textbf{Neumann} boundary condition is $\dds X^i|_b=0$. In this case the ends are not fixed any more, so that in the \textbf{open channel} thermal
partition function, one needs to integrate over all momenta running in the loop,
\begin{align}
Z_{\mathcal{N}}^{[o]}(L,R)&=\sum_n \int \frac{d^{d-2}P_{T}}{(2\pi)^{d-2}}\frac{E_n(R)}{E_{n,P_{T}}(R)}\langle n,P_{T};L|n,P_{T};0\rangle \notag \\
&= \delta^{(d-2)}(0_T)\sum_n \int d^{d-2}P_{T}e^{-E^o_{n,P_{T}}(R)L}\notag\\
&=2V_{\perp}L^{2-d}\sum_n \left(\frac{E^o_n(R)L}{2\pi}\right)^{\frac{d-1}{2}} K_{\frac{d-1}{2}}(E^o_n(R)L)~,
\end{align}
where we have used again the relativistic relations \eqref{LIMeasure}, \eqref{DeltaOperator} and \eqref{Dispersion relation}
(using the Lorentz generator that is rotating $X^0$ and the transverse directions, which is not broken in this case) together with the identity
 $\delta^{(d-2)}(0_T)=\frac{V_{\perp}}{(2\pi)^{d-2}}$ and another integral formula,
\begin{align}\label{Integral formula2}
 &\int_0^{\infty} dx~x^ne^{-\sqrt{x^2+\alpha^2}}=\frac{1}{\sqrt{4\pi}}\Gamma\left(\frac{n+1}{2}\right)(2\alpha)^{\frac{n+2}{2}} K_{\frac{n+2}{2}}(\alpha)~.
\end{align}
$V_{\perp}$ is the physical volume of the transverse dimensions; since the string can be located
anywhere in these dimensions, we expect the partition function to be proportional to $V_{\perp}$.

 In the \textbf{closed channel} the \textbf{Neumann} boundary condition implies that the initial and final closed string states
  $|\tilde B\rangle$ have no transverse momentum,
 and the amplitude is (for $\tilde v_n(L) \equiv \langle n|\tilde B\rangle$)
  \begin{align}
 Z_{\mathcal{N}}^{[c]}(L,R)&=\langle \tilde B,0_T|e^{-\mathcal{H}^cR}|\tilde B,0_T\rangle\notag\\
&=\sum_n \int \frac{d^{d-2}P_{T}}{(2\pi)^{d-2}}\frac{E_{n}^c(L)}{E_{n,P_{T}}^c(L)} \langle \tilde B,0_T|n,P_T\rangle \langle n,P_T|\tilde B,0_T\rangle  e^{-E^c_{n,P_{T}}(L)R}\notag\\
 &=V_{\perp}\sum_n\abs{\tilde v_n(L)}^2 e^{-E^c_n(L)R}~.
 \end{align}
Below we use again dimensionless quantities and define also
\begin{align}
\epsilon_n \equiv E_n/\sqrt{T}~,~\mathcal{V}_{\perp}=V_{\perp}T^{\frac{d-2}{2}}~.
\end{align}
 The summary for the partition functions in all cases and channels is:
\begin{align}
&Z_{\mathcal{D}}^{[o]}(l,r)=\sum_{n} e^{-\epsilon^{o}_n(r)l}~, \label{DOCPF} \\
&Z_{\mathcal{D}}^{[c]}(l,r)=2r^{2-d}\sum_{n} f_n(l) \left(\frac{\epsilon^{c}_n(l)r}{2\pi}\right)^{\frac{d-1}{2}} K_{\frac{d-3}{2}}(\epsilon^{c}_n(l)r)~,\label{DCCPF}\\
&Z_{\mathcal{N}}^{[o]}(l,r)=2l^{2-d}\mathcal{V}_{\perp}\sum_n  \left(\frac{\epsilon^o_n(r)l}{2\pi }\right)^{\frac{d-1}{2}} K_{\frac{d-1}{2}}(\epsilon^o_n(r)l)~,\label{NOCPF}\\
&Z_{\mathcal{N}}^{[c]}(l,r)=\mathcal{V}_{\perp}\sum_n \tilde f_n(l) e^{-\epsilon^{c}_n(l)r}~,\label{NCCPF}
\end{align}
where we define $f_n(l)\equiv|v_n(L,T)|^2 T^\frac{d-2}{2}$ and $\tilde f_n(l)\equiv|\tilde v_n(L,T)|^2 T^\frac{2-d}{2}$.

In general the partition function at any order can be interpreted through its open or closed string channel. We will now use the above partition function forms
to write down their general expansions in all cases. Then we expand our explicit computations from the previous section in the same manner, and compare them
with the general expansions to extract constraints on the couplings, as well as the corrections to the energies at each level.
In the Dirichlet case we use the closed channel in order to find constraints on the couplings; they show up there since this is where we have used the
implicit parts of Lorentz. We do not obtain here the complete constraints (taking into account the
full information coming from Lorentz symmetry), due to the unknown wave functions of the boundary closed string states; a similar computation on
the torus, where there are no boundary states, was performed in \cite{Aharony:2009} and the complete constraints on the bulk couplings were indeed found.
The open channel is used in order to find the open string energy corrections, whereas the closed string energies cannot be corrected by  boundary terms,
and the corrections from bulk terms have already been calculated in \cite{Aharony:2009}. In the Neumann case the use of the implicit parts of Lorentz
is in the open channel, and thus this is where constraints will be found. Since the open channel does not include any unknown wave functions, in this case
the derived constraints are expected to be complete, as for the torus.

\subsection{The free case (level 0)}

\subsubsection{Dirichlet boundary conditions}

\textbf{At level zero} we have the partition function \cite{Dietz:1982uc}
\begin{equation}\label{PF_0}
Z_{\mathcal{D}}^{(0)}=e^{-(r+2\mu)l} \eta(q)^{2-d}~.
\end{equation}

  For the \textbf{open channel} one can expand $\eta(q)$ in powers of $q\equiv e^{-\frac{\pi l}{r}}$,
 \begin{align}
& \eta(q)^{2-d}=\sum_{n=0}^{\infty} \omega_n q^{\frac{2-d}{24}+n}~,
 \end{align}
 with:
  \begin{align}
\omega_0=1~,~\omega_1=d-2~,&~\omega_2=\frac{(d-2)(d+1)}{2}~,~\omega_3=\frac{(d-2)(d-1)(d+6)}{6}~,~\ldots
 \end{align}
  In its open channel form \eqref{DOCPF} we have\footnote{Note that we are slightly abusing notations here, where we are using the index $n$ both for the summation over all string states, and for the summation over the free action energy levels. We use $\sum_n$ for the former, and $\sum_{n=0}^{\infty}$ for the latter. }
\begin{equation}\label{Free PF Open Channel}
Z_{\mathcal{D}}^{(0)}(l,r)=\sum_{n=0}^{\infty} \omega_n e^{-\eonz(r)l}~,
\end{equation}
from which we find the open string energies,
\begin{equation}\label{Open String Energies}
\eonz(r)=r+2\mu+\frac{\pi}{r}\left(n-\frac{d-2}{24}\right)~,
\end{equation}
and the open string degeneracies $\omega_n$ of the free action.

For the \textbf{closed channel} we first write a general power series expansion of the closed string energies and wave functions,
\begin{align}
&\epsilon^c_{n}(l)=l\left( 1+\ecnb l^{-2}+\ecnd l^{-4}+...\right)~\equiv~\ecnz(l)+ l\left(\ecnd l^{-4}+\ecnf l^{-6}...\right)~, \label{ecn expansion}
\end{align}
\begin{align}
&f_n(l)\equiv\fnz\left( 1+f_{n,1}l^{-1}+ f_{n,2}l^{-2}+... \right)~,
\label{fn expansiond}
\end{align}
and then put them in the closed channel partition function (\ref{DCCPF}) and expand again to lowest order,
\begin{align}\label{DCCPFE_0}
&Z_{\mathcal{D}}^{[c]}(l,r)=\left(\frac{l}{2\pi r}\right)^{\frac{d-2}{2}}\sum_{n} \fnz e^{-\ecnz(l)r}
\left[1+O(l^{-1})\right]~ .
\end{align}
The modular transformation property of $\eta(q)$ \eqref{Modular Transformations} is now used to rewrite \eqref{PF_0} in its closed channel form,
\begin{align}\label{free closed PF}
& Z_{\mathcal{D}}^{(0)}(l,r)~=~e^{-(r+2\mu)l}\left( \frac{l}{2r} \right)^{\frac{d-2}{2}} \eta(\tilde q)^{2-d}= e^{-2\mu l} \left( \frac{l}{2r}
\right)^{\frac{d-2}{2}}\sum_{n=0}^{\infty} \omega_n e^{-\ecnz(l)r} ~,
\end{align}
with $\tilde q\equiv e^{-\frac{4\pi r}{l}}$. This gives the closed string energies
\begin{equation}\label{ecnz}
\ecnz=l+\frac{4\pi}{l}\left(n-\frac{d-2}{24}\right)~,
\end{equation}
and the boundary state wave functions (summed over all states at level $n$)
\begin{align}\label{fnzD}
\sum_{i_n=1}^{\omega_n^c}F_{ni_n}(l)=e^{-2\mu l}\pi^{\frac{d-2}{2}}\omega_n~,
\end{align}
where we have split the index $n$ going over all energy states
into an index $n$ going over the free action energy levels, and an index $i_n$ running over the $\omega_n^c$ different closed string states that are degenerate at zeroth order
in the $n$'th energy level, but (possibly) split at a higher order.

\subsubsection{Neumann boundary conditions}

In the \textbf{open channel} the partition function \eqref{NOCPF} is expanded into
\begin{align}\label{DCCPFEN_0}
&Z_{\mathcal{N}}^{[o]}(l,r)=\mathcal{V}_{\perp}\left(\frac{r}{2\pi l}\right)^{\frac{d-2}{2}}\sum_{n} e^{-lr\left(1+\frac{\eona}{r}+\frac{\eonb}{r^2}\right)}
\left[1+O(r^{-1})\right]~,
\end{align}
to which we compare the free action partition function \eqref{eqNPF0},
\begin{align}
Z_{\mathcal{N}}^{(0)}=&\mathcal{V}_{\perp}\left(\frac{r}{2\pi l}\right)^{\frac{d-2}{2}}\sum_{n=0}^{\infty} \omega_n e^{-lr\left(1+\frac{2\mu}{r}+\frac{\pi}{r^2}[n-\frac{d-2}{24}]\right)}~,
\end{align}
finding the same open string energies and degeneracies as in the Dirichlet case at level zero.

Similarly, in the \textbf{closed channel} with
\begin{align}
&\tilde f_n(l)\equiv\fnzt\left( 1+\tilde f_{n,1}l^{-1}+ \tilde f_{n,2}l^{-2}+... \right)~, \label{fn expansion}
\end{align}
 we compare the expansion of \eqref{NCCPF},
 \begin{align}
 Z_{\mathcal{N}}^{[c]}(l,r)=\mathcal{V}_{\perp}\sum_n \fnzt e^{-rl\left(1+\frac{\ecna}{l}+\frac{\ecnb}{l^2}\right)}\left[1+O(l^{-1})\right]~,
 \end{align}
 with \eqref{eqNPF0} after applying the modular transformation \eqref{Modular Transformations},
 \begin{align}
 Z_{\mathcal{N}}^{(0)}(l,r)=\left(\frac{1}{4\pi}\right)^{\frac{d-2}{2}}V_{\perp}e^{-l(r+2\mu)}\eta(\tilde q)^{2-d}~,
 \end{align}
 to find the same closed string energies as in Dirichlet at level zero, and the following wave functions (summed over all states at level $n$, with a similar split of index as before):
 \begin{align}
 \sum_{i_n=1}^{\omega_n^c}\tilde F_{ni_n}(l)=e^{-2\mu l}(4\pi)^{\frac{2-d}{2}}\omega_n~.
 \end{align}

\subsection{Higher levels}

In order to compare the higher level contributions to the partition function \eqref{eqDPF1}-\eqref{eqNPF3} with the general forms \eqref{DOCPF}-\eqref{NCCPF},
we need to expand the open channel partition functions in powers of $1/r$, and the closed channel ones in powers of $1/l$. The comparison gives constraints
on the couplings, as well as the corrections for the energy levels. For the comparison of the closed channel we first perform a modular transformation on
the partition function so that it can be manifestly expanded.

\subsubsection{Dirichlet boundary conditions}

We begin with the closed channel since this is where we will find the constraints, which we can later use also in the open channel.

In the \textbf{closed channel}, expanding all terms in (\ref{DCCPF}) while using \eqref{ecn expansion}-\eqref{fn expansiond}, \eqref{ecnz}-\eqref{fnzD}, we get (see \eqref{closed expansion derivation})
\begin{align}\label{DCCPFE}
Z_{\mathcal{D}}^{[c]}(l,r)&=\sum_{n} 2f_n(l) r^{2-d} \left(\frac{\epsilon^c_n(l)r}{2\pi}\right)^{\frac{d-1}{2}}K_{\frac{d-3}{2}}(\epsilon^c_n(l)r)~\nn \\
&=e^{-(r+2\mu)l}\left(\frac{l}{2 r}\right)^{\frac{d-2}{2}} \tilde q^{\frac{2-d}{24}} \sum_{n=0}^{\infty}\omega_n \tilde q^n
\left\{1+\left[\widehat f_{n,1}\right]\frac{1}{l}\right.+\\
  &\hphantom{X}\left.+\left[-\widehat\ecnd t^{-1} +\left(\widehat f_{n,2}+\frac{d-2}{2}\ecnb\right)+\frac{(d-2)(d-4)}{8}t\right]\frac{1}{l^2}\,+\right. \nn \\
  &\hphantom{X} \left. + \left[-\widehat {f_{n,1}\ecnd}t^{-1}+\left(\widehat f_{n,3}+ \frac{d-2}{2}\widehat f_{n,1}\ecnb\right)+\frac{(d-2)(d-4)}{8}\widehat f_{n,1}t\right]\frac{1}{l^3}+\ldots\right\}~, \nn
\end{align}
generalizing the lowest order term (\ref{DCCPFE_0}), where $t\equiv \frac{l}{r}$. In the above, getting from the first to the second line, again we have made the same split of index as before, and we also defined the following averaged quantities, weighted by the overlap of each state with the boundary state in the free theory:
\begin{align}\label{Closed averages}
\widehat{f_{n,k}}\equiv \frac{\sum_{i_n=1}^{\omega_n^c} F_{ni_n} f_{ni_n,k}}{\sum_{i_n=1}^{\omega_n^c}F_{ni_n}}~,
~\widehat{\varepsilon^c_{n,k}}\equiv \frac{\sum_{i_n=1}^{\omega_n^c} F_{n i_n} \varepsilon^c_{ni_n,k}}{\sum_{i_n=1}^{\omega_n^c}F_{ni_n}}~,
~\widehat{\fna\varepsilon^c_{n,k}}\equiv \frac{\sum_{i_n=1}^{\omega_n^c} F_{ni_n} f_{ni_n,1}\varepsilon^c_{ni_n,k}}{\sum_{i_n=1}^{\omega_n^c}F_{ni_n}}~.
\end{align}
At each level, this expansion gives a power series in $\tilde q$, and we can compare it with the actual contributions at that level,
order by order in $\tilde q$.

\textbf{At level $1$}, applying the modular transformation \eqref{Modular Transformations} to \eqref{eqDPF1} and comparing it with \eqref{DCCPFE}
gives \cite{Luscher:2004ib} (see \eqref{Compare DC1} for more details)
\begin{align}
b_1=\widehat{\fna(...)}=0~.
\end{align}
 We can then rewrite the general expansion \eqref{DCCPFE} as
\begin{align}\label{DCCPFE1}
&Z_{\mathcal{D}}^{[c]}(l,r)=e^{-2\mu l-rl}\left(\frac{l}{2 r}\right)^{\frac{d-2}{2}} \tilde q^{\frac{2-d}{24}} \sum_{n=0}^{\infty}\omega_n \tilde q^n \times\\
&\times\left\{1+\left[-\widehat{\ecnd}t^{-1}+\widehat{\fnb}+2\pi(d-2)\left(n-\frac{d-2}{24}\right)+\frac{(d-2)(d-4)}{8}t\right]\frac{1}{l^2}
 +\left[\widehat{\fnc}\right]\frac{1}{l^3}+\ldots\right\}~,\nn
\end{align}
where we have also replaced $\ecnb$ with its value \eqref{ecnz}.

 \textbf{At level $2$}, after applying a modular transformation to \eqref{eqDPF2}, its comparison with \eqref{DCCPFE1} gives  the following equations
 \cite{Luscher:2004ib}\footnote{Notice that the factor $4$ mismatch from L\"uscher and Weisz \cite{Luscher:2004ib} is due to the factor $4$ difference in
 the normalization of $c_2$ and $c_3$.} (see \eqref{Compare DC2} for details)
\begin{align}\label{Const1}
&(d-2)c_2+c_3=\frac{d-4}{8}~, \notag \\
&{f_{0,2}}=\frac{\pi(d-2)}{6}~, \notag \\
&{\varepsilon^c_{0,4}}=\frac{\pi^2 (d-2)}{18}[2c_2+(d-1)c_3]~.
\end{align}
 A similar calculation on the torus \cite{Aharony:2009} gave the full constraints (in dimensionless couplings)\footnote{Notice that this is what one obtains
 by assuming that $c_2$ and $c_3$ are $d$-independent; this is not a coincidence, and is understood when computing the constraints directly
 by demanding the invariance under the implicit part of Lorentz \cite{Aharony:2010}, as we explain below.},
\begin{align}\label{Const_2}
&c_2=\frac{1}{8}~~,~~c_3=-\frac{1}{4}~~~,
\end{align}
and then we can write:
\begin{align}\label{Const_22}
\varepsilon^c_{0,4}=-8\pi^2\left(\frac{d-2}{24}\right)^2~,~~f_{0,2}=\frac{\pi(d-2)}{6}~.
\end{align}
The corrections to higher energy levels of the closed string are similarly extracted by comparing terms with higher powers of $\tilde q$
(alternatively, by using relations (\ref{series1}) and going through some algebra, all energy levels are extracted at once):
\begin{align}\label{Correct_2}
&\widehat{\ecnd}=-8\pi^2\left(n-\frac{d-2}{24}\right)^2~,~~\widehat{\fnb}=-4\pi\left(n-\frac{d-2}{24}\right)~.
\end{align}
 Our correction to the wave function \eqref{Correct_2} at this level corrects a small error in \cite{Aharony:2009}.
 The obtained couplings are exactly those obtained from a long string expansion of the Nambu-Goto string, and so are the energy corrections.

\textbf{At level $3$}, comparing \eqref{DCCPFE1} with \eqref{eqDPF3}, we find after applying modular transformations that
 \begin{align}\label{Const_3}
 b_3=0~,
 \end{align}
  and no constraint is found for $b_2$ (see \eqref{Compare DC3} for the details).
The correction to the ground state wave function is
\begin{align}
\foc=\frac{4b_2\pi^3(d-2)}{15}~.
\end{align}
The closed string energies are not corrected at this level since we had here only boundary contributions, so the only corrections are to the closed string
boundary state wave functions. Higher order terms in $\tilde q$ give the corrections to the higher wave functions.

In the  \textbf{open channel}, string energies can be expanded in $1/r$,
\begin{align}\label{OCEE}
 \epsilon^o_n(r)=r(1+\eona r^{-1} + \eonb r^{-2}+ ...)= \eonz(r)+r(\eonc r^{-3}+\eond r^{-4}+...)~.
\end{align}
Then, making the same split of index as before with a similar definition of averages,
\begin{align}\label{Open averages}
\widehat{\varepsilon^o_{n,k}}\equiv\frac{1}{\omega_n}\sum_{i_n=1}^{\omega_n}\varepsilon^o_{n,i_n,k}~,
~\widehat{(\varepsilon^o_{n,k})^2}\equiv\frac{1}{\omega_n}\sum_{i_n=1}^{\omega_n}(\varepsilon^o_{n,i_n,k})^2~,...
\end{align}
 we find the general expansion,
\begin{align}
Z_{\mathcal{D}}^{[o]}(l,r)&=\sum_{n=0}^{\infty}\sum_{i_n=1}^{\omega_n} e^{-\epsilon^o_{n,i_n}(r)l}\notag \\
 &=\sum_{n=0}^{\infty}\omega_n  e^{-\eonz(r)l}\left\{ 1- \left[\widehat{\eonc}\right]\frac{l}{r^2}-\left[\widehat{\eond}-\half
   \widehat{(\eonc)^2t}\right] \frac{l}{r^3}\right. - \notag \\
 &\hspace{3.5cm}\left.-\left[\widehat{\eone}-\widehat{\eonc\eond}t+\frac{1}{6}\widehat{(\eonc)^3}t^2\right]\frac{l}{r^4} -...\right\} ~.
\end{align}
As for the closed expansion, we compare the contributions from the general open expansion and from the corresponding computation results at a specific level.
We will already use here the constrained values for the couplings $c_2=\frac{1}{8},c_3=-\frac{1}{4}$ and $b_1=b_3=0$.
Since $b_1=0$ we have $\widehat{\eonc(...)}=0$, and the expansion is simpler:
\begin{align}\label{DOCPFE}
&Z_{\mathcal{D}}^{[o]}(l,r)=\sum_{n=0}^{\infty}\omega_n  e^{-\eonz(r)l}\left\{ 1-\left[\widehat{\eond}\right]\frac{l}{r^3}- \left[\widehat{\eone}\right]\frac{l}{r^4} -...\right\} ~.
\end{align}

\textbf{At level $2$}, comparing (\ref{eqDPF2}) with (\ref{DOCPFE}) we obtain
 \begin{align}
\widehat{\eond}=-\frac{\pi^2}{2}\left(n-\frac{d-2}{24}\right)^2~
 \end{align}
  for the corrections to the open string energies at this level (for details see \eqref{Compare DO2}).

\textbf{At level $3$}, we compare the order $O(q^0)$ terms in (\ref{eqDPF3}) and (\ref{DOCPFE}) to find the corresponding correction to the ground-state energy
(see details in \eqref{Compare DO3}),
\begin{align}\label{}
\varepsilon^o_{0,5}=-\frac{b_2\pi^3(d-2)}{60}~.
 \end{align}
Higher energy level corrections are similarly extracted by comparing higher powers of $q$ (see footnote \ref{lowlevels} in appendix \ref{appendixE}).

\subsubsection{Neumann boundary conditions}

In the \textbf{open channel}, the energy expansion (\ref{OCEE}) is put into the partition function (\ref{NOCPF}) which is then expanded in powers of $1/r$,
\begin{align}
Z_{\mathcal{N}}^{[o]}(l,r)&=2l^{2-d}\mathcal{V}_{\perp}\sum_{n}  \left(\frac{\epsilon^o_n(r)l}{2\pi }\right)^{\frac{d-1}{2}} K_{\frac{d-1}{2}}(\epsilon^o_n(r)l)\\
&=\mathcal{V}_{\perp}\left(\frac{r}{2\pi l}\right)^{\frac{d-2}{2}}e^{-l(r+2\mu)}q^{\frac{2-d}{24}}\sum_{n=0}^{\infty}
  \omega_n q^n\left\{1+\left[\frac{d-2}{2}\eona-\widehat\eonc t\right]\frac{1}{r}\right.+\nn\\
&\hphantom{X}\left.+\left[\frac{d(d-2)}{8}t^{-1}+\frac{d-2}{2}\eonb-\widehat \eond t+\mu^2\#_n\right]\frac{1}{r^2}+\left[-\widehat\eone t+\mu\#_n+\mu^3\#_n\right]\frac{1}{r^3}+...\right\}~,\nn
\end{align}
where we have kept terms proportional to powers of $\mu$ in an implicit form, and defined similar averages as in previous cases.
  We define $\tilde Z\equiv \mathcal{V}_{\perp}\left(\frac{r}{2\pi l}\right)^{\frac{d-2}{2}}e^{-l(r+2\mu)}q^{\frac{2-d}{24}}$,
  and use the explicit values for $\eona~,\eonb$\,, and get
\begin{align}
&\qquad  Z_{\mathcal{N}}^{[o]}(l,r)=\tilde Z \sum_{n=0}^{\infty} \omega_n q^n \left\{1+\left[\mu(d-2)-\widehat\eonc t\right]\frac{1}{r}\right.+\\
&\left.+\left[\frac{d(d-2)}{8}t^{-1}+\frac{\pi(d-2)(n-\frac{d-2}{24})}{2}-\widehat \eond t +\mu^2\#_n\right]\frac{1}{r^2}+\left[-\widehat\eone t
  +\mu\#_n+\mu^3\#_n\right]\frac{1}{r^3}+...\right\}~.\nn
\end{align}
This expansion is to be compared with the explicit partition function computations \eqref{eqNPF0}-\eqref{eqNPF3},
\begin{align}
&Z_{\mathcal{N}}(l,r)=\tilde Z \left(\sum_{n=0}^{\infty} \omega_n q^n\right)\left\{1+2a_1(d-2)\left[1-\frac{\pi}{12}E_2(q)t\right]\frac{1}{r}\right.+\nn\\
&\left.+\left[\left\{-d(d-2)(c_2+c_3)t^{-1}+\frac{\pi(d-2)}{12}[2c_2+(d-1)c_3]E_2(q)\right.\right.\right.+\nn\\
&\left.\left.\left.\hspace{0.8cm}+\pi^2\left([2c_2+(d-1)c_3]\frac{2}{24^2}E_2(q)^2+[2(d-1)c_2+
(d+1)c_3]H_{2,2}(q)\right)t\right\}+\left\{\half\langle S_1'^2\rangle\right\}\right]\frac{1}{r^2}\right.+\nn\\
&\left.+\left[\left\{-\frac{a_2\pi^3(d-2)}{60}E_4(q)\right\}
+\left\{-2a_3d(d-2)t^{-1}+\frac{a_3\pi}{3}d(d-2)E_2(q)+\frac{a_3d}{6}E_2(q)^2 t+\langle S'_1S_2\rangle\right\}\right.\right.-\nn\\
&\left.\left.\hspace{0.6cm}-\left\{\frac{1}{6}\langle S_1'^3\rangle\right\}\right]\frac{1}{r^3}+...\right\}~,
\end{align}
where we have separated different powers of $\mu$ (knowing a posteriori how the couplings depend on $\mu$) inside different powers of $t$ inside
different powers of $1/r$. We point out that in the Neumann case, since we will show that the level $1$ coefficient is non-vanishing ($a_1=\mu/2$),
at higher levels the computation is more involved, and the complete comparison and the extraction of constraints demands the computation of multi-vertex functions
 such as $\langle S_1'^2\rangle$ and others, which are harder to compute. We refrain from computing those here, and instead use at higher levels our knowledge of the constraints
 on the coefficients ($a_3=-\mu/8$, $a_2$ is unconstrained) from the direct method that is explained in the next section. Note that the
 $a_1$ and $a_3$ terms are just part of the Neumann Nambu-Goto (NNG) action, which includes the additional Nambu-Goto-like term on the boundary in static gauge
\begin{align}
S'^{NG}_{\mathcal{N}}=\mu\boundint \sqrt{1+\ddz X\cdot \ddz X}~.
\end{align}
 This is just the free particle action for the massive end-points.

From the comparison at \textbf{level $1$} we find the constraint $a_1=\frac{\mu}{2}$, and the correction to the energies
\begin{align}
\widehat\eonc=-2\pi\mu\left(n-\frac{d-2}{24}\right)~.
\end{align}

 At \textbf{level $2$} we obtain the complete constraints $c_2=\frac{1}{8}~,~c_3=-\frac{1}{4}$, and the energy level corrections,
  \begin{align}
  \widehat\eond=-\half \left(n-\frac{d-2}{24}\right)^2+\mbox{(a possible $\mu^2$-term from the NNG action)}~,
   \end{align}
   where the possible $\mu^2-$correction comes from $\langle S_1'^2\rangle$ (if it is not canceled by the $\mu^2-$terms in the general expansion).

 At \textbf{level $3$} we see that the energy correction splits into the contribution from the free parameter $a_2$, which contributes to the ground state energy
 \begin{align}
 \varepsilon^{o,a_2}_{0,5}=\frac{a_2\pi^3(d-2)}{60}~,
 \end{align}
  and other possible NNG contributions proportional to $\mu$ and $\mu^3$.

The \textbf{Neumann string in the closed channel} is not really interesting here since it does not lead to constraints, nor any correction to the energy levels
(since the closed string energies cannot be corrected by any additional boundary action). The only new information encoded in this channel is the corrections
to the wave functions, and we will not bother to compute them here, although they can easily be extracted in a similar manner to the previous cases.

 \section{Direct constraints from Lorentz invariance}\label{Section5}

 The effective action we use manifests the invariance under $SO(d-2)$ rotations of the coordinates orthogonal to the worldsheet $X^i$ ($i=2,...,d-1$). In addition, in the bulk of the worldsheet it also manifestly preserves the $SO(1,1)$ invariance of the $X^0-X^1$ plane. The complete $SO(1,d-1)$ Lorentz invariance is spontaneously broken by the classical solution around which we expand, but the expanded action should still respect this symmetry non-linearly\footnote{The gauge fixing $\dda X^{\beta}=\delta_{\alpha}^{\beta}$ with Dirichlet (Neumann) boundary conditions $\ddt X^i=0$ ($\dds X^i=0$) explicitly break the Lorentz generators $\Sigma_{0i}$ ($\Sigma_{1i}$), and in that case only $\Sigma_{1i}$ ($\Sigma_{0i}$) should be required as a symmetry.}. To derive the form of the symmetry transformations in the static gauge, consider,
 for example, a rotation in the $X^1-X^2$ plane
 \begin{align}
 \delta_{12} X^1=\epsilon X^2~~,~~\delta_{12} X^2=-\epsilon X^1~.
 \end{align}
 In order to keep the gauge fixing $\sigma^1=X^1$ we must then also make a diffeomorphism
\begin{align}
\delta_{12} \sigma^1=\epsilon X^2(\sigma)~~,~~ \delta_{12} \sigma^0=0~.
\end{align}
The complete Lorentz transformation law in the static gauge, that should leave
 the action invariant, is then \cite{Aharony:2010}
\begin{align}\label{delta12}
\delta_{12}(\dda X^i)=-\epsilon\delta_{\alpha 1}\delta^{i2}-\epsilon\dda ( X^2\dds X^i)~.
\end{align}
Similarly, the rotation of $X^0$ with $X^2$ induces
\begin{align}\label{delta02}
\delta_{02}(\dda X^i)=-\epsilon\delta_{\alpha 0}\delta^{i2}-\epsilon\dda ( X^2\ddt X^i)~.
\end{align}
It is easy to verify that operating with these symmetry transformations on the bulk action and demanding the result to vanish, the previous constraints are obtained \cite{Aharony:2010}
\begin{align}
c_2=\frac{1}{8}~,~c_3=-\frac{1}{4}~.
\end{align}
Similarly, operating with $\delta_{12}$ on the Dirichlet boundary action and demanding the result to vanish gives
\begin{align}
b_1=b_3=0~,
\end{align}
while doing the same with $\delta_{02}$ on the Neumann boundary action gives
\begin{align}
a_1=\frac{\mu}{2}~,~a_3=-\frac{\mu}{8}~.
\end{align}
 In fact, while the constraints for the ratios between the different couplings in each case are indeed obtained directly by only using the
 local symmetry transformations \eqref{delta12}-\eqref{delta02}, the overall normalization is fixed by global properties, and is slightly more subtle.
We refer the reader to \cite{Aharony:2010} for further details.

\section{An example : open strings in holographic confining gauge theories}\label{Section6}

Our discussion above is general and applies to any effective action around a long open string in flat space. In particular,
we can consider a specific realization of such a scenario, in a confining gauge theory that has a weakly coupled and weakly curved dual string theory description, such as \cite{Witten:1998zw}\nocite{Maldacena:2000yy}-\cite{Klebanov:2000hb}.
We consider a long fundamental superstring confined to some IR region and stretched between two D-branes\footnote{We ignore the dynamics of the
D-branes themselves since they must have infinite extent for the string to be able to end
on them.}, and we should find an effective action of the form we wrote above.
An interesting question is whether there are more constraints, or whether the effective action around the confined long string will have the allowed
free parameters turned on. We will verify below that the allowed boundary terms discussed in the previous sections are indeed turned on, and that no disallowed terms are generated.

We consider specifically a type II superstring in a class of confining backgrounds discussed
in \cite{Aharony:2009}, in which there is a minimal radial coordinate and a cycle that vanishes smoothly
at that value.  The expanded Euclidean action (in inverse powers of the tension), including the necessary terms for the integration out of the
massive modes at \textbf{1-loop}\footnote{For all the details and notations for this action, its derivation and related information we refer the reader
to section 4 of \cite{Aharony:2009} where a detailed presentation is given.} is \cite{Aharony:2009},
\begin{align}\label{Original action}
S_{boson}= &\ T\int d^2\sigma\left\{\left(1+\frac{1}{2T}\dda X\cdot \ddau X\right)\left(1+\frac{1}{2T} \ddb Y \cdot \ddbu Y + \frac{1}{2T} m_b^2 Y_b^2\right)\right.\nn\\
& \qquad \qquad\qquad\qquad \left.-\frac{1}{2T^2}\dda X\cdot \ddb X \ddau Y\cdot \ddbu Y+...\right\} ~.
\end{align}
$X^i~(i=2,..,d-1)$ are the flat coordinates of the effective string which are massless fields on the worldsheet.
 $Y_b~(b=1,...,N_B)$ are the
coordinates in additional curved directions corresponding to massive fields on the worldsheet, which are integrated out to give the effective string action in flat space.
In addition to ignoring terms with higher powers of $X$'s (that are irrelevant to our analysis) and terms with higher powers of $Y$'s
(that do not contribute at 1-loop), we neglect here completely the fermions and some other fields (the intrinsic metric, the kappa-symmetry-fixing ghosts
and possible additional perturbatively massless coordinates); we will claim below that these do not change our final conclusions.
We use the conventions $X\cdot X=\sum_i X^i X^i~,~Y\cdot Y=\sum_b Y_b Y_b$. The action (\ref{Original action}) was obtained \cite{Aharony:2009} by considering the kappa fixed
Green-Schwarz superstring action in the mentioned family of confining backgrounds, expanded in the number of heavy fields and keeping only the operators
that include up to two of them, needed for 1-loop order. Each loop order comes with a power of $\frac{m^2}{T}$, where $m$ is the mass of some heavy mode,
which is small by construction in the weakly curved background we consider (this is necessary so that the worldsheet theory is weakly coupled and under control).
We then integrate out the massive modes to get the effective action for the massless ones, which generically will include contributions to all possible
operators in the bulk and on the boundary\footnote{For example,
 1-loop corrections to quadratic terms are of the form
$\sim\frac{m^2}{T}\int d^2\sigma \partial^2 X^2~F(\frac{\delta}{m},\frac{\partial}{m})$, where $\delta$ stands for the Dirac delta function,
that will result with an operator on the boundary. An operator of the form $\sim\int d\sigma \partial^4 X^2$, that we are expecting,
is obtained with $F=\frac{\delta \partial^2}{m^3}$, and so its coupling is of order $\frac{1}{Tm}$. We will see this explicitly below.}.

 \subsection{Integrating out the heavy modes}

For simplicity, we choose in our computation of this section the worldsheet to be the half-plane
 $\mathds{R}^2_+\equiv\{(\sigma^0,\sigma^1)|\sigma^0\in \mathds{R},\sigma^1\in \mathds{R}_+\}$;
 since the action is local, for the purpose of computing the effective action, as long as we have a boundary, the choice of worldsheet is a matter of convenience.
We begin with Dirichlet boundary conditions for both the $X$ and $Y$ transverse coordinates.
Integration out is carried out by
\begin{align}\label{Integrating out}
&e^{-S_{Eff}[X]}\equiv\int DY e^{-S[X,Y]}= \\
&=\exp\left[-\int d^2\sigma\left(T+\half\dda X\cdot \ddau X\right)\right] \int DY \exp\left[  -\half \int d^2\sigma Y_b (-\partial^2+ m_b^2) Y_b\right]\times\nn\\
&\hphantom{X} \times\left\{1-\frac{1}{2T}\int d^2\sigma\dda X\cdot \ddb X\left[\half\delta^{\alpha \beta}\left(\partial_{\gamma} Y\cdot \partial^{\gamma} Y +
m_b^2 Y_b^2\right)-\ddau Y\cdot \ddbu Y\right]+...\right\} ~.\nn
\end{align}

Integrating over the $X$-independent term (for details see appendix \ref{integrating}) gives the bosonic part of the correction to the constant terms in the bulk and on the boundary,
 \begin{align}\label{IO_0}
&I_0\equiv \int DY e^{-\half \int d^2\sigma Y_b (-\partial^2+ m_b^2) Y_b}=\exp\left[-\Delta T_B\int_{\mathds{R}_+^2} d^2\sigma -\mu_B\int_{\mathds{R}}d\sigma^0\right]~,
\end{align}
where $\Delta T_B$ corrects the string tension
 \begin{align}
&\Delta T_B=-\frac{1}{8\pi}\sum_b m_b^2\log (m_b^2)+\mbox{divergences~,}
\end{align}
and
\begin{align}
\mu_B=-\frac{1}{8}\sum_b m_b+\mbox{divergences~.}
\end{align}
The divergences must and do cancel, when combining the contribution from the massive fermions (that we have neglected).
The total correction to the string tension (independent of having any boundary) was found \cite{Aharony:2009,Bertoldi:2004rn} before to
be\footnote{We recall that the finiteness of this correction (there is an implicit cut-off dependence in the $log$s) requires the general identity,
 \begin{align}
 \sum_b m_b^2=\sum_f m_f^2~,
 \end{align} which is assumed to be valid in any action of this kind.}
\begin{align}\label{Tension Correction}
\Delta T\equiv \Delta T_B+\Delta T_F=\frac{1}{8\pi}\left(\sum_f m_f^2\log (m_f^2)-\sum_b m_b^2\log (m_b^2)\right)~.
\end{align}

Integrating out the mixed operators as well gives (for details see \eqref{IO_2;0}),
\begin{align}
 I_2&\equiv \frac{1}{2T}\int_{\mathds{R}_+^2}d^2\sigma \dda X\cdot \ddb X ~\langle\half \delta^{\alpha\beta}\left(\partial_{\gamma}Y\cdot
     \partial^{\gamma}Y+m_b^2Y_b^2\right)-\ddau Y\cdot \ddbu Y\rangle\nn\\
 &=\frac{\Delta T_B}{T}\int_{\mathds{R}_+^2} \half \dda X\cdot \ddau X+b_2^B\int_{\mathds{R}}d\sigma^0 \ddt\dds X\cdot \ddt\dds X
     ~\line(0,+1){9}\line(0,-1){10}_{\{\sigma^1=0\}}+...~,
\end{align}
with the bosonic contribution to the boundary coupling
\begin{align}
b_2^B=-\frac{1}{64T}\sum_b \frac{1}{m_b}~,
\end{align}
where the ellipsis stands for higher derivative terms on the boundary.
 Up to the derivative order we work in, and up to 1-loop order (in $m^2/T$) and second order in $X$, the resulting effective action is
\begin{align}
S_{eff}=\int_{\mathds{R}_+^2}d^2\sigma \left[T'+\half\dda X'\cdot \ddau X'\right]+\int_{\mathds{R}}d\sigma^0 \left[\mu_B+b_2^B\ddt\dds X'\cdot \ddt\dds X'\right]
~\line(0,+1){9}\line(0,-1){10}_{\{\sigma^1=0\}} ~,
\end{align}
with the corrected tension $T'=T+\Delta T_B$ and the field wave-function renormalization $X'=X(1+\frac{\Delta T_B}{2T})$
(there are also the fermionic contributions that we have ignored).

 Some clarifications are in place. First, the original theory is understood to be finite (it is manifestly so in a different gauge)
 and divergences are expected to cancel out after taking into account all neglected fields, and so we just ignore divergent terms in our case.
  The other fields will leave also finite contributions.
  We do not expect, however, that the contributions from other fields could cancel a non-zero contribution to $b_2$ of the kind that we find, in the general case,   since their contributions are independent of the masses $m_b$;
  the fermions can contribute  with a similar scale $m_f$, but no generic cancelation is possible since our contribution goes as $\sum_b (1/m_b)$
  and the only sum rule is $\sum_b m_b^2-\sum_f m_f^2=0$.
   We thus see that the $b_2$ coupling generically shows up in this framework, and we see its dependence on the bosonic masses,
  though it will also get some finite contribution from the fermions (and possibly other fields) as well.
   We see also that the $b_1$ coupling does not get contributions from the bosonic heavy sector,
   and as a result of our analysis in the previous sections we are assured that it remains zero also after the inclusion of all other fields.

The case with Dirichlet boundary conditions for both the $X$'s and all the $Y$'s corresponds to the case when the long string is stretched at the minimal radial position,
between two $D0$-branes; however, this is not possible by charge conservation. For a realistic scenario we need to consider some other boundary conditions for at least some of the fields.
 When allowing also for Neumann boundary conditions for the $Y$'s, the computation is very similar, and the result is that the contributions to $\mu$ or $b_2$ only flip their sign for each Neumann $Y$-field. For a general combination of several orthogonal directions ($Y^{a'},a'=1,..,p$) with Neumann boundary conditions, and several orthogonal directions
    ($Y^{a},a=1,...,N_B-p$) with Dirichlet boundary conditions, the resulting $\mu$ and $b_2$ are (considering only their bosonic contribution)
     \begin{align}\label{general b_2}
 \mu_B=-\frac{1}{8}\left[\sum_a m_a-\sum_{a'} m_{a'}\right]~,~
b_2^B=-\frac{1}{64T}\left[\sum_{a}\frac{1}{m_{a}}-\sum_{a'}\frac{1}{m_{a'}}\right]~.
 \end{align}
This case correspond to a $Dp$-brane, which is localized in $\mathds{R}^d$ but stretched in some other directions (necessarily including the radial direction).

Next, consider Neumann boundary conditions for the $X$'s. This computation is similar as well (see \eqref{IO_2;3}), and the resulting effective action,
including only terms with up to two $X$'s and four derivatives on the boundary, is
 \begin{align}
S_{eff}&=\int_{\mathds{R}_+^2} d^2\sigma\left[T'+\half\dda X'\cdot \ddau X'\right]\,+\nn\\
&\hspace{1cm}+\int_{\mathds{R}}d\sigma^0\left[\mu_B+a_1^B \ddt X'\cdot \ddt X'+a_2^B \ddt^2 X'\cdot \ddt^2 X'\right]\line(0,+1){9}\line(0,-1){8}_{\{\sigma^1=0\}}~,
\end{align}
with
 \begin{align}
 a_1^B=-\frac{1}{16T}\left[\sum_{a} {m_{a}}-\sum_{a'} {m_{a'}}\right]~,~a_2^B=-\frac{1}{64T}\left[\sum_{a}\frac{1}{m_{a}}-\sum_{a'}\frac{1}{m_{a'}}\right]~,
 \end{align}
where we have used in the above the operator identity  $\ddt^2 X\cdot (\ddt^2+\dds^2) X=0$. Of course $\mu$ and $\Delta T$ do not change from the previous case. Notice that the computed effective couplings on the boundary (in units of the string tension) obey the expected relation from Lorentz symmetry,
\begin{align}
&a_1^B=\frac{1}{2}\mu^B~.
\end{align}
%
We do not consider in this paper mixed Dirichlet and Neumann boundary conditions among the $X$'s, but a similar
  analysis can be easily done for that case as well. We see that in all analyzed cases, there is a perfect matching between the expected effective action,
  and the computed effective action in the holographic framework, and also that all couplings that are not constrained by the Lorentz symmetry
   indeed show up.
  Finally, we note that when the worldsheet theory has
for each Dirichlet mode a corresponding Neumann mode with the same mass, all our 1-loop couplings vanish. It can be easily checked that in that (and only that) case,
all quadratic terms on the boundary, at any derivative order, vanish in the effective theory.

\acknowledgments
We would like to thank N. Klinghoffer, Z. Komargodski and A. Schwimmer for many interesting
discussions and for collaborations on related topics, and to thank B. Bringoltz, J. Sonnenschein,
M. Teper and V. Vyas for useful discussions. OA would like to thank ECT*, Trento for hospitality during the conclusion of this work, and the participants in the ``Confining flux tubes and strings'' workshop there for useful discussions; this was supported in
part by the European Community - Research Infrastructure Action under the
FP7 ``Capacities'' Specific Programme, project ``HadronPhysics2''.
This work was supported in part by the Israel--U.S.~Binational Science Foundation, by a research center supported by the Israel Science Foundation (grant number 1468/06), by a grant (DIP H52) of the German Israel Project Cooperation, and by the Minerva foundation with funding from the Federal German Ministry for Education and Research.

\appendix
\section{Functions and their modular transformations}
\label{appendixA}

Below are some functions that appear often in our partition function calculations. The notation for their variables is the following:
\begin{align}
&\tau\equiv i\frac{l}{2r}~,& q\equiv e^{2\pi i \tau}=e^{-\frac{\pi l}{r}} ~, \notag \\
&\tilde\tau \equiv -\frac{1}{\tau}=i\frac{2r}{l}~,&\tilde{q}\equiv e^{2\pi i \tilde\tau}=e^{-\frac{4 \pi r}{l}} ~.
\end{align}
The Dedekind-$\eta$-function is
\begin{align}
\eta(q)\equiv q^{\frac{1}{24}}\prod_{n=1}^{\infty}(1-q^n)~.
\end{align}
The Eisenstein series and their derivative are:
\begin{align}\label{series1}
E_{2k}(q)&\equiv  1+\frac{2}{\zeta(1-2k)}\sumn \frac{n^{2k-1}q^n}{1-q^n}~,\notag\\
H_{2,2k}(q)&\equiv \frac{\zeta(1-2k)}{2}q\frac{d}{d q}E_{2k}(q)=\sumn \frac{n^{2k}q^n}{(1-q^n)^2}~.
\end{align}
Specifically,
\begin{align}
&E_2(q)=24q\frac{d}{dq}\log\eta(q)~,\notag \\
 &H_{2,2}(q)=-\frac{1}{24}q\frac{d}{dq}E_2(q)= \frac{E_4(q)-E_2(q)^2}{288}~.
\end{align}
We use the following expansions,
\begin{align}
&E_2(q)=1-24q-3\cdot 24 q^2-4\cdot 24 q^3-7\cdot 24 q^4-... \notag \\
&E_2(q)^2=1-2\cdot 24q+18\cdot 24 q^2+136\cdot 24 q^3+202\cdot 24 q^4-... \notag \\
&E_4(q)=1+10\cdot 24q+90\cdot 24 q^2+... \notag \\
&H_{2,2}(q)=q+6q^2+12q^3+28q^4+...
\end{align}

The functions defined above all have simple transformation properties under the modular transformation $\tau\rightarrow -\frac{1}{\tau}$:
\begin{align}\label{Modular Transformations}
\eta(q)&=(-i\tilde\tau)^{1/2}\eta(\tilde q)=\left(\frac{2r}{l}\right)^{\half}\eta(\tilde q)~, \notag \\
E_2(q)&=-\frac{6i}{\pi}\tilde \tau+\tilde \tau^2 E_2(\tilde q)= \frac{12r}{\pi l}-\left(\frac{2r}{l}\right)^2 E_2(\tilde q)=\frac{12r}{\pi l}\left(1-\frac{\pi r}{3l}E_2(\tilde q)\right)~, \notag\\
H_{2,2}(q)&=\frac{\log(\tilde q)^2}{4\pi^4}\left[-\frac{1}{8}-\frac{1}{48}\log(\tilde q)E_2(\tilde q)+\frac{1}{4}\log(\tilde q)^2 H_{2,2}(\tilde q)\right] ~& \notag \\ &=\frac{\tilde \tau^2}{8 \pi^2}\left[1+\frac{\pi i \tilde \tau}{3} E_2(\tilde q)+8\pi^2\tilde \tau^2 H_{2,2}(\tilde q)\right]~& \notag\\
&=-\half\left(\frac{r}{\pi l}\right)^2\left[1-\frac{2\pi r}{3l}E_2(\tilde q)-2\left(\frac{4\pi r}{l}\right)^2H_{2,2}(\tilde q)\right]~, \notag \\
E_4(q)&=\tilde \tau^4 E_4(\tilde q)=\left(\frac{2r}{l}\right)^4 E_4(\tilde q)~.
\end{align}

For regularization the $\zeta$-function is used
\begin{align}
\zeta(s)\equiv \sumn n^{-s}~,
\end{align}
and specific values that appear in the calculations are:
\begin{align}\label{}
 &\zeta(0)=-\half ~,~\zeta(-1)=-\frac{1}{12}~,~\zeta(-3)=\frac{1}{120}~,\nn\\
 &\zeta(-2n)=0~~\forall~ n\in\mathds{N}~,~\zeta'(0)=-\half\log(2\pi)~.
\end{align}

\section{Regularization of sums}

In our computations we encounter one finite sum,
\begin{align}
\summ \frac{1}{\denom} = \frac{\pi rl}{2n} \coth\argc,
\end{align}
and we use the\textbf{ $\zeta$-function regularization} to regularize the others:
\begin{align}\label{sums}
&\summ m^k =0~,~~~\forall~k=0,1,2,... \nn \\
&\sumn n^s \coth \argc =\zeta(-s) + 2\sumn \frac{n^s q^n}{1-q^n}= \zeta(-s)E_{s+1}(q)~, \nn \\
&\sumn n^s \coth^2 \argc =\zeta(-s) + 4\sumn \frac{n^s q^n}{(1-q^n)^2}= \zeta(-s)+4H_{2,s}(q)~, \nn \\
&\sum_{m,n} \frac{n^l m^{2k}}{\denom}=(-1)^k\pi r^2\left(\frac{l}{2r}\right)^{2k+1} \zeta(1-l-2k)E_{2k+l}(q) ~.
\end{align}
The details for the two middle identities can be found in \cite{Aharony:2009}.
Here is the computation for the first identity:
\begin{align}
&\summ m^0=1+2\sum_{m=1}^{\infty}m^0=1+2\zeta(0)=0~, \notag \\
&\summ m^{2n+1}=\sum_{m=1}^{\infty} m^{2n+1} + \sum_{m=1}^{\infty} (-m)^{2n+1} = 0~, \notag \\
&\summ m^{2n}=0+2\zeta(-2n)=0~.
\end{align}
The computation for the last identity uses:
\begin{align}
J_k& \equiv \summ \frac{m^{2k}}{\denom}= \summ m^{2(k-1)}\left( 1- \frac{\frac{n^2}{r^2}}{\denom}\right)\frac{l^2}{4}~ \notag \\
&=-\frac{n^2l^2}{4r^2}\summ\frac{m^{2(k-1)}}{\denom}~=~-\left(\frac{nl}{2r}\right)^2 J_{k-1}~,
\end{align}
and also
\begin{equation}J_0=\frac{\pi rl}{2n}\coth(\frac{n\pi l}{2r})~,\end{equation}
so that
  \begin{align}
  J_k=(-1)^k\frac{n^{2k-1} \pi l^{2k+1}}{2^{2k+1}r^{2k-1}}\coth\argc~.
  \end{align}
  Then,
\begin{align}
\sum_{m,n} \frac{n^l m^{2k}}{\denom}&=\sumn n^l J_k ~=~(-1)^k\pi r^2\left(\frac{l}{2r}\right)^{2k+1}\sumn n^{l+2k-1}\coth\argc~ \notag \\
&=(-1)^k\pi r^2 \left(\frac{l}{2r}\right)^{2k+1}\left\{\zeta(-l-2k+1)+2\sumn \frac{n^{l+2k-1}q^n}{1-q^n}\right\}~ \notag \\
&= (-1)^k \pi r^2\left(\frac{l}{2r}\right)^{2k+1} \zeta(1-l-2k)E_{2k+l}(q)~.
\end{align}
In particular,
\begin{align}
&\sum_{m,n} \frac{m^{2}}{\denom}=-\pi r^2\left(\frac{l}{2r}\right)^{3} \zeta(-1)E_{2}(q)=\frac{\pi l^3}{96 r}E_2(q)~, \label{SUMm2} \\
&\sum_{m,n} \frac{n^2}{\denom}=\pi r^2\left(\frac{l}{2r}\right) \zeta(-1)E_{2}(q)=-\frac{\pi rl}{24}E_2(q)~, \label{SUMn2} \\
&\sum_{m,n} \frac{m^{4}}{\denom}=\pi r^2\left(\frac{l}{2r}\right)^{5} \zeta(-3)E_{4}(q)=\frac{\pi l^5}{(60\cdot 64)r^3}E_4(q)~,\label{SUMm4}\\
&\sum_{m,n} \frac{n^2 m^{2}}{\denom}=-\pi r^2\left(\frac{l}{2r}\right)^{3} \zeta(-3)E_{4}(q)=-\frac{\pi l^3}{(60\cdot 16)r}E_{4}(q)~,\label{SUMm2n2}\\
&\sum_{m,n} \frac{n^4}{\denom}=\pi r^2\left(\frac{l}{2r}\right)\zeta(-3)E_{4}(q)=\frac{\pi rl}{(60\cdot 4 )}E_{4}(q)~.\label{SUMn4}
\end{align}

\section{Propagators}

The Green's function on a two dimensional manifold $\Sigma$ is defined as a propagator through
\begin{align}
G(\sigma,\sigma')\delta^{ij}\equiv \langle X^i(\sigma) X^j(\sigma')\rangle = \frac{\int \mathcal{D}X e^{-S_0[X]}X^i(\sigma)X^j(\sigma')}{\int \mathcal{D}X e^{-S_0[X]}}~,
   \end{align}
   where $\sigma\in \Sigma$ and $X^i$ are scalar bosons with the free action $S_0=\sum_i \half\int_{\Sigma}d^2 \sigma X^i (-\partial^2) X^i$.

   A direct computation of the free path integral is obtained by expanding the $X^i$'s with a \textbf{complete} and \textbf{orthonormal} set
    of eigenfunctions of the Laplacian operator on $\Sigma$,
    \begin{align}
    -\partial^2 \Psi_I(\sigma)=\lambda_I \Psi_I(\sigma)~.
    \end{align}
    The result of this computation is
  \begin{align}
   G(\sigma,\sigma')=\sum_{I\neq 0}\frac{\Psi_I^*(\sigma) \Psi_I(\sigma')}{\lambda_I}~,
   \end{align}
with the summation not including the zero mode $\Psi_0$ (if present).
   Equivalently the Green's function is also defined through
\begin{align}
-\partial_{\sigma}^2 G(\sigma,\sigma')=\delta(\sigma-\sigma')-|\Psi_0|^2~,
\end{align}
solved trivially by the same form above. Notice that the definition of a manifold with a boundary includes the choice for the boundary conditions.
In this paper we need the propagator on the cylinder (with period $l$ and length $r$) and on the half plane, with Dirichlet or Neumann boundary conditions. For completeness we also present the details for the propagator on the plane.

\subsection{On the cylinder}

 In the Dirichlet case the complete orthonormal set of eigenfunctions of the Laplacian operator is
\begin{align}\label{DL_eigenfunctions}
\left\{\Psi_{m,n}(\sigma)=\sqrt{\frac{2}{rl}}\sin\argn \exp\argm\right\}_{(m,n)_{\mathcal{D}}}~,
\end{align}
where $(m,n)_{\mathcal{D}}\in \mathds{Z}\times \mathds{N}$,
with the resulting propagator,
\begin{align}\label{calDprop}
G_{\mathcal{D}}(\sigma,\sigma')=\frac{2}{\pi^2rl}\sum_{(m,n)_D} \frac{\sin\argo\sin\argop\exp\left(\frac{2\pi im(\sigma^0-\sigma'^0)}{l}\right)}{\denom}~.
\end{align}

In the case of Neumann boundary conditions the complete orthonormal set is
\begin{align}\label{NL_eigenfunctions}
\hspace{-1.5cm}\left\{\Psi_{m,n}(\sigma)=\sqrt{\frac{2}{rl}}\cos\argn \exp\argm\right\}_{(m,n)_{\mathcal{D}}}\bigcup ~ \left\{\Psi_{m,0}(\sigma)=\sqrt{\frac{1}{rl}}\exp\argm\right\}_{m \in\mathds{Z}}~.
\end{align}
Notice the different normalization for the $n=0$ modes that is given by orthonormality, and that in this case we also have a zero mode ($n=m=0$).
The resulting propagator is
\begin{align}\label{calNprop}
\hspace{-1.3cm}G_{\mathcal{N}}(\sigma,\sigma')=\frac{2}{\pi^2rl}\left(\sum_{(m,n)_D} \frac{\cos\argo\cos\left(\frac{n\pi\sigma'^1}{r}\right) \exp\left(\frac{2\pi im(\sigma^0-\sigma'^0)}{l}\right)}{\denom}+\frac{l^2}{8}\sum_{m\neq 0}\frac{\exp\left(\frac{2\pi im(\sigma^0-\sigma'^0)}{l}\right)}{m^2}\right)~.
\end{align}

\subsection{On the plane}

On the plane the complete orthonormal set is
\begin{align}
\left\{\Psi_{k}(\sigma)=\frac{1}{2\pi}e^{ik\cdot \sigma}\right\}~,
\end{align}
where both $\sigma$ and $k$ are defined on the plane.
The resulting propagator is
\begin{align}
G(\sigma,\sigma')=\int_{\mathds{R}^2}\frac{d^2 k}{(2\pi)^2}\frac{e^{ik\cdot(\sigma-\sigma')}}{k^2+m^2}~,
\end{align}
where we allow for a general mass $m$ for the scalar. For the massless scalar the propagator diverges and needs to be IR-regulated.
In momentum (Euclidean) space,
\begin{align}
X(\sigma)\equiv\frac{1}{2\pi}\int_{\mathds{R}^2}d^2k~e^{ik\cdot\sigma}\tilde X(k)~,
\end{align}
where $\tilde X(-k)=\tilde X^*(k)$ for the reality of $X(\sigma)$, the propagator is
\begin{align}
\tilde G(k,k')\equiv \langle \tilde X(k) \tilde X(k')\rangle =\frac{1}{k^2+m^2}\delta(k+k')~.
\end{align}

\subsection{On the half plane}

On the half plane and in the case of Dirichlet boundary conditions, the complete orthonormal set is
\begin{align}
\left\{\Psi_{k}(\sigma)=\frac{i}{\pi}\sin(k_1\sigma^1)e^{ik_0\sigma^0}=\frac{1}{2\pi}\left[e^{ik\cdot\sigma}-e^{ik\cdot\bar\sigma}\right]\right\} ~,
\end{align}
where both $\sigma$ and $k$ are defined on the upper half plane, and the resulting propagator is
\begin{align}
G_{\mathcal{D}}(\sigma,\sigma')&=\int_{\mathds{R}^2_+}\frac{d^2 k}{\pi^2}\frac{\sin(k_1\sigma^1)\sin(k_1\sigma'^1)e^{ik_0(\sigma^0-\sigma'^0)}}{k^2+m^2} \\
&=\int_{\mathds{R}^2}\frac{d^2 k}{(2\pi)^2}\frac{\left[e^{ik\cdot(\sigma-\sigma')}-e^{ik\cdot(\sigma-\bar{\sigma'})}\right]}{k^2+m^2}
=G_{\mathds{R}^2}(\sigma,\sigma')-G_{\mathds{R}^2}(\sigma,\bar{\sigma'})~, \nn
\end{align}
where we define $\bar \sigma \equiv (\sigma^0,-\sigma^1)$.
In momentum space
\begin{align}\label{half plane momentum space}
X(\sigma)\equiv\frac{1}{2\pi}\int_{\mathds{R}_+^2}d^2k~\left[e^{ik\cdot\sigma}-e^{i \bar k\cdot\sigma}\right]\tilde X(k)~,
\end{align}
 where $\tilde X(k)$ is defined only on the half-plane, and obeys $\tilde X^*(k)=-\tilde X(-\bar k)$ for the reality of $X(\sigma)$, and the propagator is
\begin{align}\label{MSpropD}
\mathcal{}\tilde G_{\mathcal{D}}(k,k') =-\frac{1}{k^2+m^2}\delta(k+\bar k')~.
\end{align}

Similarly, in the case of Neumann boundary conditions the complete orthonormal set is
\begin{align}
\left\{\Psi_{k}(\sigma)=\frac{1}{\pi}\cos(k_1\sigma^1)e^{ik_0\sigma^0}=\frac{1}{2\pi}\left[e^{ik\cdot\sigma}+e^{ik\cdot\bar\sigma}\right]\right\}
~~~\mbox{for $k_1>0$} ~,
\end{align}
and
\begin{align}\label{special}
\left\{\Psi_{k}(\sigma)=\frac{1}{\sqrt{2}\pi}e^{ik_0\sigma^0}\right\} ~~~\mbox{for $k_1=0$} ~.
\end{align}
We define the momentum decomposition to include this normalization, so that $\int_{\mathds{R}_+^2}d^2 k$ includes a relative factor of half on the line $k_1=0$,
\begin{align}\label{}
X(\sigma)\equiv\frac{1}{2\pi}\int_{\mathds{R}_+^2}d^2k~\left[e^{ik\cdot\sigma}+e^{i \bar k\cdot\sigma}\right]\tilde X(k)~.
\end{align}
With this the resulting propagator in momentum space
differs from the Dirichlet case only by a minus sign,
\begin{align}
G_{\mathcal{N}}(\sigma,\sigma')=G_{\mathds{R}^2}(\sigma,\sigma')+G_{\mathds{R}^2}(\sigma,\bar{\sigma'})~.
\end{align}
In momentum space,
\begin{align}
\tilde G_{\mathcal{N}}(k,k')=+\frac{1}{k^2+m^2}\delta(k+\bar k')~.
\end{align}

\section{Ignoring operators proportional to the free equation of motion}
\label{appendixD}

In this appendix we argue that operators proportional to the free action equations of motion do not contribute to the cylinder partition function, so we can ignore them in our analysis.
An operator that is proportional to the free e.o.m. or its derivatives is of the general form $\partial \cdots \partial \partial^2 X \cdot \partial \cdots \partial X$ and
it can contribute in the partition function generally through
\begin{align}
\partial \cdots \partial \partial^2  \partial' \cdots \partial' G(\sigma,\sigma') ~.
\end{align}
Since $\partial^2 G(\sigma,\sigma') =-\delta(\sigma-\sigma')$ (up to a possible constant), this contribution is proportional to some derivatives of a delta function and so
if $\sigma\neq\sigma'$ it identically vanishes. The Laplacian operator ($\partial^2$) exactly cancels the denominator in the propagator \eqref{calDprop},
and in the case when $\sigma=\sigma'$ this contribution is then proportional to $\summ m^k$, and thus identically vanishes as well, under our regularization
(see \eqref{sums}). For the Neumann case \eqref{calNprop}, the additional contribution $\sum_{m\neq 0} m^k$ also vanishes, except when $k=0$, and then it equals
 $1$, but the case of $k=0$ never appears\footnote{It actually could have appeared from a term in the bulk $\bulkint \partial^2 X\cdot X$,
 that is invariant under translations in the Neumann case. However, with the Neumann boundary conditions, this term is identically equivalent to the
 free term, and thus need not be written.}. This formal identity, $\partial^2\equiv 0$, holds also on the boundary.

\section{Detailed calculations}
\label{appendixE}
\subsection{The partition function}

\textbf{Dirichlet, level 0:}\\
The (Euclidean) Laplacian on the cylinder with Dirichlet boundary condition has the complete orthonormal set of eigenfunctions given above \eqref{DL_eigenfunctions},
  with the corresponding eigenvalues $\lambda_{n,m}=-(\frac{\pi n }{r})^2-(\frac{2\pi m}{l})^2$, giving the determinant:
\begin{align}
\det(-\partial^2)=\prod_{(m,n)_{\mathcal{D}}}\left[\left(\frac{\pi n}{r}\right)^2+\left(\frac{2\pi m}{l}\right)^2\right]~.
\end{align}
The partition function is then:
\begin{align}\label{calDPF0}
Z_{\mathcal{D}}^{(0)}&=\int DX e^{-S_E}=e^{-l(r+2\mu)}\int DX e^{-\half\int d^2\sigma X^i (-\partial^2) X^i} \notag\\ &=e^{-l(r+2\mu)}\det\left(-\frac{2\pi}{\partial^2}\right)^{\frac{d-2}{2}}=e^{-l(r+2\mu)}\exp\left(\frac{d-2}{2}\sum_{(m,n)_{\mathcal{D}}} \log\left(\frac{2/\pi}{\frac{n^2}{r^2}+\frac{4m^2}{l^2}}\right)\right)\notag\\
&\equiv e^{-l(r+2\mu)}\exp\left(\frac{d-2}{2} f(r,l)\right).
\end{align}
We see that\footnote{We use here $\summn\equiv\summnD$.}
\begin{align}
\partial_r f(r,l)&=\frac{2}{r^3}\summn \frac{n^2}{\frac{n^2}{r^2}+\frac{4m^2}{l^2}}=  \frac{2}{r^3} \pi r^2\left(\frac{l}{2r}\right) \zeta(-1)E_{2}(q)~\notag \\
&=-\frac{\pi l}{12r^2}E_2(q)=-2\frac{\pi l}{r^2}q\frac{\partial}{\partial q}\log(\eta(q))=-\frac{\partial}{\partial r}2\log(\eta(q))~,
\end{align}
where we have used the $\zeta$-function regularization for the summation of the infinite sum plus other identities and conventions all elaborated thoroughly in the appendix.
Similarly,
\begin{equation}
\partial_l f(r,l)=-\frac{\partial}{\partial l}2\log(\eta(q))~,
\end{equation}
so that we find
\begin{equation}
f(r,l)=-2\log(\eta(q))~,
\end{equation}
up to a constant which we argue to equal zero (so that it is compatible with the partition function). Thus,
\begin{align}
Z_{\mathcal{D}}^{(0)}= e^{-l(r+2\mu)}\eta(q)^{2-d}~.
\end{align}

\textbf{Dirichlet, level 1:}\\
With the propagator \eqref{calDprop} we compute (using \eqref{SUMn2}),
\begin{align}\label{calDPF1}
 \langle S'_1 \rangle  = b_1 \boundint \langle\dds X^i \dds X^i\rangle =  b_1 (d-2) \boundint \dds \ddsp G_{\mathcal{D}}~,
 \end{align}
 \begin{align}
\boundint \dds \ddsp G_{\mathcal{D}} &=\frac{4}{r^3} \summn  \frac{n^2}{\denom}=-\frac{\pi l}{6 r^2}E_2(q)  ~.
\end{align}

\textbf{Dirichlet, level 2:}\\
We use the following convention,
\begin{align}
 \dda \ddb' G \equiv \lim_{\sigma\rightarrow \sigma'} \partial_{\sigma^{\alpha}} \partial_{\sigma'^{\beta}} G(\sigma,\sigma')~,
\end{align}
and then compute (as in \cite{Luscher:2004ib,Aharony:2009})
\begin{align}\label{calDPF2}
\langle S_2 \rangle &=c_2 \bulkint \langle \dda X^i \ddau X^i ~\ddb X^j \ddbu X^j \rangle +c_3\bulkint \langle \dda X^i \ddb X^i~\ddau X^j \ddbu X^j \rangle \notag \\
&=~ \{[(d-2)^2c_2+(d-2)c_3]I_1+[2(d-2)c_2+(d-2)^2c_3+(d-2)c_3]I_2\}~\notag\\
&=~ (d-2)\{[(d-2)c_2+c_3]I_1+[2c_2+(d-1)c_3]I_2\}~,
\end{align}
where:
\begin{align}\label{}
 &I_1=\bulkint \dda \ddaup  G_{\mathcal{D}}~ \ddb \ddbup G_{\mathcal{D}}=\bulkint\left\{(\ddz\ddz'G_{\mathcal{D}})^2+(\ddo\ddo'G_{\mathcal{D}})^2+2\ddz\ddz'G_{\mathcal{D}}\ddo\ddo'G_{\mathcal{D}}\right\}~, \nn\\
 &I_2=\bulkint \dda\ddbp G_{\mathcal{D}} ~\ddau \ddbup G_{\mathcal{D}}=\bulkint\left\{(\ddz\ddz'G_{\mathcal{D}})^2+(\ddo\ddo'G_{\mathcal{D}})^2+2(\ddz\ddo' G_{\mathcal{D}})^2\right\}~.
 \end{align}
\begin{align}
&\bulkint (\ddz\ddz'G_{\mathcal{D}})^2 =\nn\\
&=\frac{64}{r^2 l^5}\summn\summnp \frac{m^2}{\denom}\frac{m'^2}{\denomp}\underbrace{\int d\sigma^1 \sin^2\argo\sin^2\argop}_{=\frac{r}{8}(2+\delta_{n,n'})}\nn\\
&=\frac{8}{rl^5}\left(2\left(\summn\frac{m^2}{\denom}\right)^2+\sum_n \left(\sum_m\frac{m^2}{\denom}\right)^2\right)\nn\\
&=\frac{8}{rl^5}\left(2\left(\frac{\pi l^3}{96 r}E_2(q)\right)^2+\left(\frac{\pi l^3}{8r}\right)^2\sum_n n^2 \coth^2\argc\right)\nn \\
&=\frac{\pi^2 l}{8r^3}\left(\frac{1}{72}E_2(q)^2+4H_{2,2}(q)\right)=\frac{\pi^2 l}{576r^3}E_4(q)~.
\end{align}
\begin{align}
&\bulkint (\ddo\ddo'G_{\mathcal{D}})^2 =\nn\\
&=\frac{4}{r^6l}\summn\summnp \frac{n^2}{\denom}\frac{n'^2}{\denomp}\underbrace{\int d\sigma^1 \cos^2\argo\cos^2\argop}_{=\frac{r}{8}(2+\delta_{n,n'})}\notag\\
&=\frac{1}{2r^5l}\left(2\left(\summn\frac{n^2}{\denom}\right)^2+\sum_n n^4\sum_m\frac{1}{\denom}\sum_{m'}\frac{1}{\left(\frac{n^2}{r^2}+\frac{4m'^2}{l^2}\right)}\right)\notag\\
&=\frac{1}{2r^5l}\left(2\left(-\frac{\pi rl}{24}E_2(q)\right)^2+\left(\frac{\pi rl}{2}\right)^2\sum_n n^2 \coth^2\argc\right)\notag \\
&=\frac{\pi^2 l}{8r^3}\left(\frac{1}{72}E_2(q)^2+4H_{2,2}(q)\right)=\frac{\pi^2 l}{576r^3}E_4(q)~.
\end{align}
\begin{align}
&\bulkint \ddz\ddz'G_{\mathcal{D}}\ddo\ddo'G_{\mathcal{D}}=\nn\\ &=\frac{16}{r^4l^4}\summn\frac{m^2}{\denom}\summnp \frac{n'^2}{\denomp}
 \underbrace{\bulkint\cos^2\argo\sin^2\argop}_{=\frac{rl}{8}(2-\delta_{n,n'})} \notag\\
&=\frac{2}{r^3l^3}\left(2\summn\frac{m^2}{\denom}\summnp\frac{n'^2}{\denomp
}-\sum_n n^2\sum_m\frac{m^2}{\denom}\sum_{m'}\frac{1}{\frac{n^2}{r^2}+\frac{4m'^2}{l^2}}\right)\notag\\
&=\frac{2}{r^3l^3}\left(-2\frac{\pi^2 l^4}{24\cdot96}(E_2(q))^2+\frac{\pi^2 l^4}{16}\sum_n n^2 \coth^2\argc \right)\notag \\
&=\frac{\pi^2 l}{8r^3}\left(-\frac{1}{72}E_2(q)^2+4H_{2,2}(q)\right)~.\\\nn\\
&\bulkint (\ddz\ddo'G_{\mathcal{D}})^2 \propto\left(\summn\frac{m}{\denom}\right)^2=0~.
\end{align}
Then \cite{Aharony:2009},
\begin{align}
&I_1=\bulkint\left\{(\ddz\ddz'G_{\mathcal{D}})^2+(\ddo\ddo'G_{\mathcal{D}})^2+2\ddz\ddz'G_{\mathcal{D}}\ddo\ddo'G_{\mathcal{D}}\right\}=\frac{2\pi^2 l}{r^3}H_{2,2}(q)~, \notag \\
 &I_2=\bulkint\left\{(\ddz\ddz'G_{\mathcal{D}})^2+(\ddo\ddo'G_{\mathcal{D}})^2+2(\ddz\ddo' G_{\mathcal{D}})^2\right\}=\frac{\pi^2 l}{288r^3}E_4(q).
\end{align}

\textbf{Dirichlet, level 3:}\\
\begin{align}\label{calDPF3}
\langle S'_3 \rangle&= b_2 \boundint \langle (\dds X\cdot \dds X ) (\dds X\cdot \dds X) \rangle + b_3 \boundint \langle \ddt\dds X\cdot \dds\ddt X \rangle~ \notag \\
&=b_2[(d-2)^2+2(d-2)]I_3+b_3(d-2)I_4~=~ (d-2)[db_2I_3+b_3I_4],
\end{align}
\begin{align}
I_3&=\boundint (\dds \ddsp G_{\mathcal{D}})^2=  ~ \notag \\
&=\frac{4}{r^6l^2} \summn \frac{n^2}{\denom} \summnp \frac{n'^2}{\denomp} \int_0^l d\sigma^0 \cos^2\argo\cos^2\argop  \line(0,+1){15}\line(0,-1){21}_{\{\sigma^1=0\}+\{\sigma^1=r\}} \notag \\
& =\frac{8}{ r^6 l}\left(\summn \frac{n^2}{\frac{n^2}{r^2}+\frac{4m^2}{l^2}}\right)^2=\frac{\pi^2}{72}\frac{ l}{ r^4} E_2(q)^2~, \\
I_4&=\boundint \ddt \dds \ddtp \ddsp G_{\mathcal{D}} =\frac{16\pi^2}{r^3l^2}\summn \frac{n^2m^2}{\denom}
=-\frac{\pi^3l}{60r^4} E_4(q)~.
\end{align}
\newpage
\textbf{Neumann at level 0:}\\
The (Euclidean) Laplacian on the cylinder with Neumann boundary condition has the complete orthonormal set of eigenfunctions \eqref{NL_eigenfunctions}.
The change from the Dirichlet case is in replacing sines with cosines, and also including the non-trivial corresponding functions with $n=0$, with appropriate normalization. Similar arguments give the \emph{primed} determinant:
\begin{align}
\det{'}(-\partial^2)=\prod_{(m,n)_{\mathcal{N}}}\left[\left(\frac{\pi n}{r}\right)^2+\left(\frac{2\pi m}{l}\right)^2\right]~,
\end{align}
where $(m,n)_{\mathcal{N}}\in\{m\in\mathds{Z},n\in\mathds{N} ~\bigcup~ m\neq 0, n=0\}$.
The partition function is then:
\begin{align}\label{calNPF0}
Z_0&=\int DX e^{-S_E}=e^{-l(r+2\mu)}\int DX e^{-\half\int d^2\sigma X^i (-\partial^2) X^i} \notag\\ &=(rl)^{\frac{d-2}{2}}\mathcal{V}_{\perp}e^{-l(r+2\mu)}\det{'}\left(-\frac{2\pi}{\partial^2}\right)^{\frac{d-2}{2}}\nn \\
&=(rl)^{\frac{d-2}{2}}\mathcal{V}_{\perp}e^{-l(r+2\mu)}\exp\left(\frac{d-2}{2}\summnN \log\left(\frac{2/\pi}{\denom}\right)\right)\notag\\
&=(rl)^{\frac{d-2}{2}}\mathcal{V}_{\perp}e^{-l(r+2\mu)}\exp\left(\frac{d-2}{2}\large[f(r,l)+g(r,l)\large]\right)\nn \\
&=\left(\frac{r}{2\pi l}\right)^{\frac{d-2}{2}}\mathcal{V}_{\perp}e^{-l(r+2\mu)}\eta(q)^{2-d}~,
\end{align}
where
\begin{align}
&f(r,l)=\summnD \log\left(\frac{2/\pi}{\denom}\right)=-2\log(\eta(q))~, \notag\\
&g(r,l)=\sum_{m\neq 0} \log\left(\frac{2l^2}{4\pi m^2}\right)=2\zeta(0)\log\left(\frac{2l^2}{4\pi}\right)+4\zeta'(0)=-\log(2\pi l^2)~,
\end{align}
and $\mathcal{V}_{\perp}$ is the volume of the (dimensionless) transverse space of the $X^i$'s coming from integrating over the zero modes. The $(rl)^{\frac{d-2}{2}}$ factor comes from the fact that the zero modes are related to the coordinates by $X^i=\chi^i_{0,0}\frac{1}{\sqrt{rl}}+(\sigma\mbox{-dependent})$.\\

\textbf{Neumann, level 1:}\\
With \eqref{calNprop} we compute,
\begin{align}\label{calNPF1}
 &\langle S'_1 \rangle =a_1\boundint \langle \ddt X\cdot \ddt X\rangle=  a_1 (d-2) \boundint \ddt \ddtp G_{\mathcal{N}}~,
 \end{align}
 \begin{align}
 \boundint \ddt \ddtp G_{\mathcal{N}} =\frac{16}{rl^2}\left\{\summnD \frac{m^2}{\denom}+ \half\summN \frac{l^2}{4}\right\} =\frac{\pi l}{6r^2}\left\{E_2(q)-\frac{12r}{\pi l}\right\}~.
\end{align}

\textbf{Neumann, level 2:}\\
The integrals in \eqref{calDPF2} are now:
\begin{align}\label{calNPF2}
 &I_1=\bulkint \dda \ddaup  G_{\mathcal{N}}~ \ddb \ddbup G_{\mathcal{N}}=\bulkint\left\{(\ddz\ddz'G_{\mathcal{N}})^2+(\ddo\ddo'G_{\mathcal{N}})^2+2\ddz\ddz'G_{\mathcal{N}}\ddo\ddo'G_{\mathcal{N}}\right\}~, \notag \\
 &I_2=\bulkint \dda\ddbp G_{\mathcal{N}} ~\ddau \ddbup G_{\mathcal{N}}=\bulkint\left\{(\ddz\ddz'G_{\mathcal{N}})^2+(\ddo\ddo'G_{\mathcal{N}})^2+2(\ddz\ddo' G_{\mathcal{N}})^2\right\}~,
\end{align}
with:
\begin{flalign}
&\ddz\ddz'G_{\mathcal{N}}=\frac{8}{rl^3}\left\{\summnD\frac{m^2\cos^2\argo}{\denom}-\frac{l^2}{8}\right\}~,\nn\\
&\ddo\ddo'G_{\mathcal{N}}=\frac{2}{r^3l}\summnD\frac{n^2\sin^2\argo}{\denom}~,\nn\\
&\ddz\ddo'G_{\mathcal{N}}=0~.
\end{flalign}
\begin{align}
&\bulkint (\ddz\ddz'G_{\mathcal{N}})^2=\\
&=\frac{64}{r^2l^5}\left\{\summnD\frac{m^2}{\denom}\summnpD\frac{m'^2}{\denom}\int d\sigma^1\cos^2\argo\cos^2\argop\right.-\nn\\
&\hspace{1.6cm}\left.-~\frac{l^2}{4}\summnD\frac{m^2}{\denom}\into\cos^2\argo+\frac{rl^4}{64}\right\}=\mbox{result}|_D-\frac{\pi}{12r^2}E_2(q)+\frac{1}{rl}~.\nn\\ \nn\\
&\bulkint (\ddo\ddo'G_{\mathcal{N}})^2=\\
&=\frac{4}{r^6l}\summnD\frac{n^2}{\denom}\summnpD\frac{n'^2}{\denom}\int d\sigma^1\sin^2\argo\sin^2\argop=\mbox{result}|_D~.\nn\\ \nn\\
&\bulkint \ddz\ddz'G_{\mathcal{N}}\ddo\ddo'G_{\mathcal{N}}= \\
&=\frac{16}{r^4l^3}\left\{\summnD\frac{m^2}{\denom}\summnpD\frac{n'^2}{\denomp}\int d\sigma^1\cos^2\argo\sin^2\argop\right.-\overline{}\nn\\
&\hspace{1.5cm}\left.-~\frac{l^2}{8}\summnD\frac{n^2}{\denom}\into\sin^2\argo\right\}=\mbox{result}|_D+\frac{\pi}{24r^2}E_2(q)~.\nn\\\nn\\
 &\bulkint (\ddz\ddo'G_{\mathcal{N}})^2 =0~.
\end{align}
\newpage
Then,
\begin{align}
I_1&=\bulkint\left\{(\ddz\ddz'G_{\mathcal{N}})^2+(\ddo\ddo'G_{\mathcal{N}})^2+2\ddz\ddz'G\ddo\ddo'G_{\mathcal{N}}\right\}\nn\\
&=\mbox{result}|_D+\frac{1}{rl}\nn=\frac{2\pi^2 l}{r^3}H_{2,2}(q)+\frac{1}{rl}~, \notag \\
I_2&=\bulkint\left\{(\ddz\ddz'G_{\mathcal{N}})^2+(\ddo\ddo'G_{\mathcal{N}})^2+2(\ddz\ddo' G_{\mathcal{N}})^2\right\}\nn\\
&=\mbox{result}|D-\frac{\pi}{12r^2}E_2(q)+\frac{1}{rl}=\frac{\pi^2 l}{288r^3}E_4(q)-\frac{\pi}{12r^2}E_2(q)+\frac{1}{rl}~.
\end{align}

\textbf{Neumann, level 3:}\\
The correction at this level is
\begin{align}\label{calNPF3}
\langle S'_3 \rangle&=a_2\boundint \langle \ddt^2  X^i \ddt^2 X^i \rangle +a_3\boundint \langle \ddt X^i \ddt X^i \ddt X^j \ddt X^j\rangle \nn\\
&=(d-2)[a_2 I_3+a_3 d I_4]~,
\end{align}
\begin{align}
 I_3&=\boundint \ddt^2 \partial_0^{'2} G_{\mathcal{N}} \nn\\
 &=\frac{64\pi^2}{rl^4}\left\{\summnD\frac{m^4}{\denom}+\frac{l^2}{8}\summN m^2\right\}\nn\\
 &=\frac{64\pi^2}{rl^4}\left\{\frac{\pi l^5}{(60\cdot 64)r^3}E_4(q)+\frac{l^2}{4}\zeta(-2)\right\}=\frac{\pi^3l}{60r^4} E_4(q)~.\\
\nn\\
  I_4&=  \boundint (\ddt \ddtp G_{\mathcal{N}})^2\nn\\
  &=\frac{128}{r^2l^5}\left(\summnD\frac{m^2}{\denom}+\frac{l^2}{8}\summN1\right)^2\nn\\
  &=\frac{128}{r^2l^5}\left(\frac{\pi l^3}{96 r}E_2(q)-\frac{l^2}{8}\right)^2=\frac{\pi^2l}{72r^4}E_2(q)^2-\frac{\pi}{3r^3}E_2(q)+\frac{2}{r^2l}~.
 \end{align}

\subsection{Expanding the general forms for the partition function}

\textbf{The closed channel with Dirichlet boundary conditions:}\\
The Bessel functions can be expanded at large arguments as
\begin{align}\label{Bessel Expansion 0}
&~K_{\frac{d-3}{2}}(x^{-1})=\sqrt{\frac{\pi x}{2}}e^{-\frac{1}{x}}\left[1+\frac{(d-2)(d-4)}{8}x+...\right]~.
\end{align}
Then, together with the expansion of the closed string energies and wave functions \eqref{ecn expansion},\eqref{fn expansiond}, and their zeroth order value \eqref{ecnz},\eqref{fnzD}, the partition function in the closed channel is
\begin{align}\label{closed expansion derivation}
Z_{\mathcal{D}}^{[c]}(l,r)&=\sum_n 2f_{n}(l) r^{2-d} \left(\frac{\epsilon^c_{n}(l)r}{2\pi}\right)^{\frac{d-1}{2}}K_{\frac{d-3}{2}}(\epsilon^c_{n}(l)r)~\notag \\
&=\sum_{n=0}^{\infty} \sum_{i_n=1}^{\omega_{n}^c} f_{ni_n}(l) \left(\frac{\epsilon^c_{ni_n}(l)}{2\pi r}\right)^{\frac{d-2}{2}}e^{-\epsilon^c_{ni_n}(l)r} \left[1+\frac{(d-2)(d-4)}{8\epsilon^c_{ni_n}(l)r}+...\right]~\notag \\
&=\sum_{n=0}^{\infty} e^{-\epsilon^c_{0,n}(l)r}\sum_{i_n=1}^{\omega_n^c} F_{ni_n}(l)\left(1+ f_{ni_n,1}l^{-1}+...\right)\left(\frac{l}{2\pi r}\right)^{\frac{d-2}{2}} \left(1+\ecnbi l^{-2} +...\right)^{\frac{d-2}{2}}\times\notag \\
 &\hphantom{X}\times\left(1-\frac{r}{l^3}( \ecndi + \ecnfi l^{-2}+...)+\half \left(\frac{r}{l^3}\right)^2( \ecndi+ \ecnfi l^{-2}+...)^2+... \right)\times\nn \\
& \hphantom{X}\times\left[1+\frac{(d-2)(d-4)}{8rl}\left( 1 +\ecnbi l^{-2}+... \right)^{-1}+...~ \right]~\notag \\
&=e^{-(r+2\mu)l}\left(\frac{l}{2 r}\right)^{\frac{d-2}{2}} \tilde q^{\frac{2-d}{24}} \sum_{n=0}^{\infty}\omega_n\tilde q^n \times \\
&\hphantom{X}\times\left\{1+\left[\widehat f_{n,1}\right]\frac{1}{l}+\left[-\widehat\ecnd t^{-1} +\left(\widehat f_{n,2}+\frac{d-2}{2}\ecnb\right)+\frac{(d-2)(d-4)}{8}t\right]\frac{1}{l^2}\right.+ \nn \\
& \left. \hspace{1.3cm}+ \left[-\widehat {f_{n,1}\ecnd}t^{-1}+\left(\widehat f_{n,3}+ \frac{d-2}{2}\widehat f_{n,1}\ecnb\right)+\frac{(d-2)(d-4)}{8}\widehat f_{n,1}t\right]\frac{1}{l^3}+\ldots\right\}~,\nn
\end{align}
with the averages defined in \eqref{Closed averages}, and with $t\equiv\frac{l}{r}$.\\
\textbf{The open channel with Neumann boundary conditions:}
\begin{align}
Z_{\mathcal{N}}^{[o]}(l,r)&=2l^{2-d}\mathcal{V}_{\perp}\sum_n  \left(\frac{\epsilon^o_n(r)l}{2\pi }\right)^{\frac{d-1}{2}} K_{\frac{d-1}{2}}(\epsilon^o_n(r)l) \nn\\
&=l^{2-d}\mathcal{V}_{\perp}\sum_n e^{-\epsilon^o_n(r)l}\left(\frac{\epsilon^o_n(r)l}{2\pi}\right)^{\frac{d-2}{2}}
\left\{1+\frac{d(d-2)}{8}(\epsilon^o_n(r)l)^{-1}+...\right\} \nn\\
&=\mathcal{V}_{\perp}e^{-(r+2\mu)l}\left(\frac{r}{2\pi l}\right)^{\frac{d-2}{2}}q^{\frac{2-d}{24}}\sum_{n=0}^{\infty} \omega_n q^n
\left\{1+\left[\frac{d-2}{2}\eona - \widehat\eonc t\right]\frac{1}{r} \right.+\nn\\
&\hphantom{X}+\left[\frac{d(d-2)}{8}t^{-1}+\frac{d-2}{2}\left(\eonb+\frac{d-2}{4}(\eona)^2\right)-\left(\widehat\eond+\frac{d-2}{2}\eona\widehat\eonc\right)t+\half\widehat{(\eonc)^2}t^2\right]\frac{1}{r^2}\,+\nn\\
&\hphantom{X}+\left[\frac{d(d-2)(d-4)}{16}\eona t^{-1}+\frac{(d-2)(d-4)}{8}\left(2\eona\eonb-\eonc+\frac{d-6}{6}(\eona)^3\right)\right.- \nn\\
&\hphantom{X}\hphantom{X}-\left(\widehat\eone+\frac{d-2}{2}\left(\eonb\widehat\eonc+\eona\widehat\eond+\frac{d-4}{4}(\eona)^2\widehat\eonc\right)\right)t\,+\nn\\
&\hphantom{X}\hphantom{X}+\left.\left.\left(\widehat{\eonc\eond}+\frac{d-2}{4}\eona\widehat{(\eonc)^2}\right)t^2-\frac{1}{6}\widehat{(\eonc)^3}t^3\right]\frac{1}{r^3}+...\right\}~,
\end{align}
with the averages defined in \eqref{Open averages}.
\subsection{Comparing the two results}
 \textbf{Closed channel, Dirichlet, level 1:}\\
 From (\ref{DCCPFE}) we read the general form of the first correction to the partition function in the closed channel,
\begin{align}
&Z_{\mathcal{D}}^{(1)}=e^{-(r+2\mu)l}\left(\frac{l}{2r}\right)^{\frac{d-2}{2}}\tilde q^{\frac{2-d}{24}}\left[\widehat{\foa}+\widehat{f_{1,1}}(d-2)\tilde q+...\right] \frac{1}{l} ~.
\end{align}

This is to be compared with \eqref{eqDPF1} after a modular transformation \eqref{Modular Transformations},
\begin{align}\label{Compare DC1}
 Z_{\mathcal{D}}^{(1)}&=-Z_{\mathcal{D}}^{(0)} \langle S'_1 \rangle = e^{-(r+2\mu)l} \eta(q)^{2-d}~\frac{b_1 \pi(d-2)}{6}\frac{l}{r^2}E_2(q)\nn\\
 &=e^{-(r+2\mu)l}\left( \frac{l}{2r} \right)^{\frac{d-2}{2}}\tilde q^{\frac{2-d}{24}}\sum_{n=0}^{\infty} \omega_n\tilde q^n \frac{b_1\pi (d-2) }{6}
   \frac{l}{ r^2}\left[\frac{12r}{\pi l}-\left(\frac{2r}{l}\right)^2 E_2(\tilde q)\right] \\
   &=e^{-(r+2\mu)l} \left( \frac{l}{2r} \right)^{\frac{d-2}{2}} \tilde q^{\frac{2-d}{24}}(1+(d-2) \tilde q+...)
   \frac{b_1\pi(d-2)}{6} \left[\frac{12}{\pi}t-4(1-24\tilde q-...)\right] \frac{1}{l}~.\nn
\end{align}
 Since $t=-\frac{4\pi}{\log(\tilde q)}$ is non analytic in $\tilde q$, we see that the two series cannot be matched non-trivially, and so
  $b_1=\widehat{f_{n,1}...}=0$ \cite{Luscher:2004ib}.\\

 \textbf{Closed channel, Dirichlet, level 2:}\\
  In the same manner as before we compare \eqref{DCCPFE1} with \eqref{eqDPF2},
 \begin{align}\label{Compare DC2}
 Z_{\mathcal{D}}^{(2)}&=-Z_{\mathcal{D}}^{(0)}\langle S_2 \rangle =\nn\\
  &=-e^{-2\mu l-rl}\eta(q)^{2-d}(d-2)\left\{\left[(d-2)c_2+c_3\right]\frac{2\pi^2 l}{r^3}H_{2,2}(q)+ \left[2c_2+(d-1)c_3\right]\frac{\pi^2 l}{288r^3}E_4(q)\right\}~ \notag \\
 &=e^{-2\mu l-rl} \left( \frac{l}{2r} \right)^{\frac{d-2}{2}}\tilde q^{\frac{2-d}{24}}(1+O(\tilde q))(d-2) \times\nn\\
 &\hphantom{X} \times\left\{\left[(d-2)c_2+c_3\right]\frac{\pi^2 l}{r^3} \left(\frac{r}{\pi l}\right)^2\left[1- \frac{2\pi r}{3l}(1+O(\tilde q))-2\left(\frac{4\pi r}{l}\right)^2  O(\tilde q)\right] \right.- \nn \\
 &\hspace{1cm} \left.\, -\left[2c_2+(d-1)c_3\right]\frac{\pi^2 l}{288r^3} \left(\frac{2r}{l}\right)^4 (1+O(\tilde q))\right\}~ \notag \\
 &= e^{-2\mu l} \left( \frac{l}{2r} \right)^{\frac{d-2}{2}}e^{-\epsilon^c_{0,0}(l)r}(d-2)\times\nn\\
 &\hphantom{X}\times\left\{-\frac{\pi^2}{18}\left[2c_2+(d-1)c_3\right]t^{-1} -\frac{2\pi}{3}[(d-2)c_2+c_3]+\left[(d-2)c_2+c_3\right]t +O(\tilde q)\right\}\frac{1}{l^2} \notag \\
 &\overset{!}{=}~e^{-2\mu l} \left( \frac{l}{2r} \right)^{\frac{d-2}{2}}e^{-\epsilon^c_{0,0}(l)r}(d-2)\times\nn\\
 &\hphantom{X}\times\left\{- \frac{1}{d-2}\widehat{\varepsilon^c_{0,4}}t^{-1} + \left(\frac{1}{d-2}\widehat{\fob}+\half\varepsilon^c_{0,2}\right)\frac{d-4}{8}t +O(\tilde q) \right\}\frac{1}{l^2}~.
\end{align}
This gives the following equations \cite{Luscher:2004ib}:
\begin{align}\label{const1}
&(d-2)c_2+c_3=\frac{d-4}{8}~, \notag \\
&\fob=-\frac{d-2}{2}\varepsilon^c_{0,2}-\frac{2\pi}{3}(d-2)[(d-2)c_2+c_3]=\frac{\pi (d-2)^2}{12}-\frac{\pi(d-2)(d-4)}{12}=\frac{\pi(d-2)}{6}~, \notag \\
&\epsilon^c_{0,4}=\frac{\pi^2 (d-2)}{18}[2c_2+(d-1)c_3]~.
\end{align}

\textbf{Closed channel, Dirichlet, level 3:}\\
Extracting and comparing now the $O(l^{-3})$ terms from the general expansion \eqref{DCCPFE1}, and the computed partition function \eqref{eqDPF3} (modular transformed), we find
\begin{align}\label{Compare DC3}
Z_{\mathcal{D}}^{(3)}&=-Z_{\mathcal{D}}^{(0)}\langle S'_3 \rangle = -e^{-(r+2\mu)l}\eta(q)^{2-d}\frac{\pi^2}{12}(d-2)\frac{l}{r^4}\left[\frac{db_3}{6}E_2(q)^2-\frac{b_2\pi}{5}E_4(q)\right] \nn\\
&= e^{-(r+2\mu)l}\left(\frac{l}{2r}\right)^{\frac{d-2}{2}}\eta(\tilde q)^{2-d}\frac{\pi^2(d-2)}{12}t^4\left[\frac{b_2\pi}{5}\left(\frac{2r}{l}\right)^4E_4(\tilde q)-\frac{db_3}{6}\left(\frac{12r}{\pi l}\right)^2 \left(1-\frac{\pi r}{3l}E_2(\tilde q)\right)^2\right]\frac{1}{l^3} \notag \\
&= e^{-(r+2\mu)l}\left(\frac{l}{2r}\right)^{\frac{d-2}{2}}\tilde q^{\frac{2-d}{24}}\sum_{n=0}^{\infty}\omega_n \tilde q^n (d-2) \left[\frac{4\pi^3b_2}{15}E_4(\tilde q)-\frac{db_3}{6}\left(t-\frac{\pi}{3}E_2(\tilde q)\right)^2\right]\frac{1}{l^3} \notag \\
&\overset{!}{=}e^{-(r+2\mu)l}\left(\frac{l}{2r}\right)^{\frac{d-2}{2}}\tilde q^{\frac{2-d}{24}}\sum_{n=0}^{\infty}\omega_n \tilde q^n\left[\widehat{\fnc}\right] \frac{1}{l^3}~,
\end{align}
which can be equated only for
\begin{align}
b_3=0~.
\end{align}
Then we are left with
\begin{align}
\frac{4\pi^3b_2(d-2)}{15}E_4(\tilde q)\sum_{n=0}^{\infty}\omega_n \tilde q^n  \overset{!}{=} ~\sum_{n=0}^{\infty}\omega_n \widehat{\fnc} \tilde q^n ~,
\end{align}
from we which we extract the correction to the first wave function
\begin{align}
\foc=\frac{4b_2\pi^3(d-2)}{15}~,
\end{align}
and $b_2$ is left unconstrained.  The closed string energies cannot be corrected by boundary contributions and the only corrections are to the boundary state wave function. Higher order terms (in $\tilde q$) give the corrections for higher wave functions.\\

\textbf{Open channel, Dirichlet, level 2:}\\
Comparing (\ref{eqDPF2}) with (\ref{DOCPFE}) we get
\begin{align}\label{Compare DO2}
 Z_{\mathcal{D}}^{(2)}&=-Z_{\mathcal{D}}^{(0)}\langle S_2 \rangle = -e^{-(r+2\mu)l}\eta(q)^{2-d}(d-2)\times\nn\\
 &\hphantom{X}\times\left\{[(d-2)c_2+c_3]\frac{2\pi^2 l}{r^3}H_{2,2}(q)+ [2c_2+(d-1)c_3]\frac{\pi^2 l}{r^3}\left[\frac{2}{24^2}E_2(q)^2+H_{2,2}(q)\right]\right\}~ \notag \\
 &=-e^{-(r+2\mu)l}q^{\frac{2-d}{24}}\frac{\pi^2(d-2)}{4}\frac{ l}{r^3}\sum_{n=0}^{\infty}\omega_n q^n\left\{(d-4)H_{2,2}(q)-(d-2)\left[\frac{2}{24^2}E_2(q)^2+H_{2,2}(q)\right]\right\}~ \notag \\
 &= e^{-(r+2\mu)l}q^{\frac{2-d}{24}}\frac{ l}{r^3} \sum_{n=0}^{\infty}\omega_n q^n \frac{\pi^2(d-2)}{2}\left[H_{2,2}(q)+\frac{d-2}{24^2}E_2(q)^2\right]~ \notag \\
 &\overset{!}{=} e^{-(r+2\mu)l}q^{\frac{2-d}{24}} \frac{l}{r^3}\sum_{n=0}^{\infty}\omega_n  q^n \left[-\widehat\eond\right]~,
 \end{align}
from which we find by a similar calculation to that in the closed channel the corrections to the open string energies at this level,
 \begin{align}
\widehat{\varepsilon^o_{n,4}}=-\frac{\pi^2}{2}\left(n-\frac{d-2}{24}\right)^2~.
 \end{align}

  \textbf{Open channel, Dirichlet, level 3:}\\
  Comparing now powers of $q$ in (\ref{eqDPF3}) and (\ref{DOCPFE}),
\begin{align}\label{Compare DO3}
 Z_{\mathcal{D}}^{(3)}&=-Z_{\mathcal{D}}^{(0)}\langle S'_3 \rangle = e^{-(r+2\mu)l}q^{\frac{2-d}{24}}\frac{l}{r^4}\frac{b_2\pi^3(d-2)}{60}E_4(q)\sum_{n=0}^{\infty}\omega_n  q^n~ \notag \\
 &=e^{-(r+2\mu)l}q^{\frac{2-d}{24}}\frac{l}{r^4}\sum_{n=0}^{\infty}\omega_n  q^n\left[-\widehat\eone\right]\notag \\
 &\Longrightarrow~ \varepsilon^o_{0,5}=-\frac{b_2\pi^3(d-2)}{60}~.
 \end{align}
Higher energy corrections are similarly obtained by comparing higher powers of $q$, but no nice general formula is found\footnote{\label{lowlevels} The lowest levels are
\begin{equation}
\widehat{\varepsilon^o_{1,5}}=-\frac{b_2\pi^3}{60}(d+238)~,~\widehat{\varepsilon^o_{2,5}}=-\frac{b_2\pi^3}{60}\left(\frac{d^2+479d+3358}{d+1}\right)~,
~\widehat{\varepsilon^o_{3,5}}=-\frac{b_2\pi^3}{60}\left(\frac{d^3+723d^2+12224d+12972}{(d-1)(d+6)}\right)~.
\end{equation}}.

\subsection{Integrating out the heavy modes}
\label{integrating}

We start with the original action of the $X$'s and the $Y'$s \eqref{Original action}, and explicitly perform the path integral over the $Y$'s, at one-loop \eqref{Integrating out}, to obtain the effective action of the $X$'s alone.

Let's begin with the $X$-independent terms. Working on the plane we would find
\footnote{Note that we are quite loose about the measure of the path integral and it is written up to an infinite multiplicative constant that drops out in any computation. In the second line we use~
$\displaystyle \sum_p\rightarrow V_{\mathds{R}^2} \int_{\mathds{R}^2} \frac{d^2 p}{(2\pi)^2}~,$
 where $V_{\mathds{R}^2}$ is the volume of the worldsheet on the plane, which is then turned into an integration $\int_{\mathds{R}^2} d^2\sigma$.}
\begin{align}\label{IO_0_plane}
I_0&\equiv \int DY \exp\left[  -\half \int_{\mathds{R}^2} d^2\sigma Y_b (-\partial^2+ m_b^2) Y_b\right]=\prod_b \det\left(-\partial^2+m_b^2\right)^{-\half}\\
&=\exp\left[-\frac{V_{\mathds{R}^2}}{2} \sum_b\int_{\mathds{R}^2} \frac{d^2 p}{(2\pi)^2}\log(p^2+m_b^2)\right]=\exp\left[-\frac{V_{\mathds{R}^2}}{4\pi} \sum_b\int_0^{\Lambda}dp ~ p\log(p^2+m_b^2)\right]\nn\\
&=\exp\left[-\frac{V_{\mathds{R}^2}}{8\pi}\sum_b\left((p^2+m_b^2)\log(p^2+m_b^2)-p^2\right) {\bigg |}_{p=0}^{p=\Lambda}\right]=\exp\left[-\int_{\mathds{R}^2} d^2\sigma (\Delta T_B+\mbox{div.})\right]\nn~~,
\end{align}
where $\Delta T_B=-\frac{1}{8\pi}\sum_b m_b^2\log (m_b^2)~$, and we suppress the quadratically divergent terms that are canceled by the fermionic contributions that we have ignored.
This is the bosonic part of the correction to the string tension, and the total contribution is \eqref{Tension Correction} \cite{Bertoldi:2004rn,Aharony:2009}.

On the half-plane naively we just have half of this contribution,
giving us the same correction to the tension integrated over the
half-plane. But this does not measure correctly the contributions
from zero modes with $p_1=0$; these modes exist when we take
Neumann boundary conditions for $Y_b$
but not in the Dirichlet case, and are counted ``half a time''
when we divided the plane partition function by a half. Thus,
there is an extra contribution from these modes, that has an
opposite sign in the Dirichlet case compared to the Neumann case.
Evaluating this contribution directly on the half-plane is subtle,
but we can easily compute it by considering instead the partition
function on a strip $0 \leq \sigma_1 \leq R$. Comparing the
partition function with Neumann boundary conditions on both sides
of the strip, to the one with Dirichlet boundary conditions on both
sides, the only difference between them is in the contribution of
the $p_1=0$ modes (all other modes have the same Laplacian in
both cases). Thus, the ratio between the two partition functions
is given by
\begin{align}
\prod_b \det\left(-\partial_0^2+m_b^2\right)^{-\half}
&=\exp\left[-\frac{V_{\mathds{R}}}{2} \sum_b\int_{-\Lambda}^{\Lambda} \frac{d p_0}{2\pi}\log(p_0^2+m_b^2)\right]=
\exp\left[-\frac{V_{\mathds{R}}}{2} \sum_b m_b \right],
\end{align}
up to divergent terms that we expect to cancel. We interpret
this as coming from an integration over both boundaries of the strip of
$\mu$ in the Neumann case, minus the same integration in the
Dirichlet case, so we deduce that a scalar with Neumann
boundary conditions contributes $m_b/8$ to $\mu$, and a
scalar with Dirichlet boundary conditions contributes $-(m_b/8)$.
The full contribution to $\mu$ thus takes the form
\begin{align}\label{mu}
\mu_B=-\frac{1}{8}\left[\sum_a m_a-\sum_{a'} m_{a'}\right]~,
\end{align}
where $a$ runs over the Dirichlet $Y$-fields and $a'$ runs over the Neumann $Y$-fields.

For the integration of the quadratic contribution,
\begin{align}\label{IO_2;0}
I_2\equiv \frac{1}{4T}\int_{\mathds{R}_+^2}d^2\sigma \dda X\cdot \ddb X \langle \delta^{\alpha\beta}\left(\partial_{\gamma}Y\cdot \partial^{\gamma}Y+m_b^2Y_b^2\right)-2\ddau Y\cdot \ddbu Y\rangle~,
\end{align}
we use \eqref{half plane momentum space} and a similar expression for $Y_b$, and get for the Dirichlet
case both for the $X$'s and for the $Y$'s, after some algebra,
\begin{align}\label{IO_2;1}
I_2&=-\frac{1}{16\pi^2T}\int_{\mathds{R}_+^2}d^2kd^2k'd^2pd^2p'\tilde X(k)\tilde X(k')\sum_b \frac{1}{p^2+m_b^2}\delta^{(2)}(p+\bar p')\times\nn\\
&\hphantom{X}\times \left\{\delta^{(2)}(k+k'+p+p')\left[k\cdot k'(p\cdot p'-m_b^2)-2p\cdot k p'\cdot k'\right]\right.+\nn\\
&\hspace{1cm}\left. +(k,k'\rightarrow \bar k,\bar k')+(k,p\rightarrow \bar k,\bar p)+
(k',p\rightarrow \bar k',\bar p)\right.-\nn\\
&\hspace{1cm}\left.-(k\rightarrow \bar k)-(k'\rightarrow \bar k')-(p\rightarrow \bar p)-(k,k',p\rightarrow \bar k, \bar k', \bar p)\right\}.
\end{align}
The difference in the computation of the  Neumann case for the $X$'s, is only by changing a sign (minus to plus and plus to minus) wherever
one of the $k$'s gets a bar relatively to the first term. In the Neumann case for the $Y$'s there is a change in sign whenever one of the $p$'s gets a bar.
Using then \eqref{MSpropD} and rearranging further we get
\begin{align}\label{IO_2;2}
I_2&=\frac{1}{16\pi^2T}\int_{\mathds{R}_+^2}d^2kd^2k'\tilde X(k)\tilde X(k')\sum_b \int_{\mathds{R}_+^2} d^2p\frac{1}{p^2+m_b^2}\times\nn\\
& \hphantom{X}\times \left\{ \delta(k+k')[k^2(p^2+m_b^2)-2(p\cdot \bar k)^2] + \delta(k+k'+p-\bar p)[k\cdot k'(p\cdot \bar p+m_b^2)-2p\cdot k \bar p\cdot k'] \right.+\nn\\
& \hspace{0.95cm}\left.+(k,k'\rightarrow \bar k,\bar k')-(k\rightarrow \bar k)-(k' \rightarrow \bar k')\right\}~.
\end{align}
Putting a radial cut-off ($\Lambda$) on the plane, the two integrals equal:
\begin{align}
&\int_{\mathds{R}_+^2}d^2 p \frac{k^2(p^2+m_b^2)-2(p\cdot \bar k)^2}{p^2+m_b^2}=\frac{\pi}{2} k^2m_b^2\log(\Lambda^2+m_b^2)-\frac{\pi}{2} k^2m_b^2\log(m_b^2)~,\nn\\
&\int_{\mathds{R}_+^2}d^2 p \frac{k\cdot k'(p\cdot \bar p+m_b^2)-2p\cdot k p\cdot \bar k'}{p^2+m_b^2}\delta^{(2)}(k+k'+p-\bar p)+(k,k'\rightarrow \bar k,\bar k')=\\
 &\hspace{5.5cm}=\delta(k_0+k'_0)\left[\Lambda(k_0^2 + k_1 k_1') - \pi m_b k_0^2 \left(1+\frac{(k_1+k'_1)^2}{4m_b^2}\right)^{-\half}\right]~,\nn
\end{align}
where in the second integral we have ignored terms that vanish in the $\Lambda=\infty$ limit.
Putting these into \eqref{IO_2;2} and ignoring diverging terms we get
\begin{align}\label{IO_2;3}
I_2&=\frac{-1}{16\pi T}\int_{\mathds{R}_+^2}d^2kd^2k'\tilde X(k)\tilde X(k')\left\{k^2\left[\delta^{(2)}(k+k')\mp\delta^{(2)}(\bar k+k')\right]\sum_b m_b^2 \log(m_b^2)\right.\pm\nn\\
&~~~\left. \pm k_0^2\delta(k_0+k'_0)\sum_b m_b \left[\left(1+\frac{(k_1+k'_1)^2}{4m_b^2}\right)^{-\half}\mp\left(1+\frac{(k_1-k'_1)^2}{4m_b^2}\right)^{-\half}\right]\right\}~,
\end{align}
where we include here also the results for Neumann boundary conditions for the fields. The plus in the $\pm$ sign refers to
Dirichlet boundary conditions for the $Y$'s, where the minus sign is for Neumann. The minuses in the two $\mp$ signs are for Dirichlet boundary conditions
for the $X$'s, and the pluses are for Neumann.

We are only concerned about corrections to the effective action that are up to four derivatives on the boundary, and so we expand \eqref{IO_2;3} in powers of $\frac{k^2}{m^2}$. In the Dirichlet case for the $X$'s we find
\begin{align}\label{IO_2;4}
I_2&=\frac{1}{2}\int_{\mathds{R}_+^2}d^2kd^2k'\tilde X(k)\tilde X(k')k^2\left[\delta^{(2)}(k+k')-\delta^{(2)}(\bar k+k')\right]\frac{\Delta T_B}{T}\,-\nn\\
&\hspace{1.5cm}-\frac{2}{\pi}\int_{\mathds{R}_+^2}d^2kd^2k'\tilde X(k)\tilde X(k')k_0^2k_1k'_1\delta(k_0+k'_0)b_2^B\nn\\
&=\frac{\Delta T_B}{T}\int_{\mathds{R}_+^2} \half \dda X\cdot \ddau X+b_2^B\int_{\mathds{R}}d\sigma^0 \ddt\dds X\cdot \ddt\dds X
~\line(0,+1){9}\line(0,-1){8}_{\{\sigma^1=0\}}~,
\end{align}
with:
\begin{align}
\Delta T = -\frac{1}{8\pi}\sum_b m_b^2\log(m_b^2)~~,~~b_2^B=-\frac{1}{64T}\left[\sum_{a}\frac{1}{m_{a}}-\sum_{a'} \frac{1}{m_a'}\right]~.
\end{align}
The resulting effective action is
\begin{align}\label{IO_2;5}
S_{eff}&=\int_{\mathds{R}_+^2} \left[T\left(1+\frac{\Delta T}{T}\right)+\half\dda X\cdot \ddau X\left(1+\frac{\Delta T}{T}\right)\right]+
\int_{\mathds{R}}d\sigma^0 \left(\mu_B+b_2^B\ddt\dds X\cdot \ddt\dds X \right)\line(0,+1){9}\line(0,-1){8}_{\{\sigma^1=0\}} \nn\\
&=\int_{\mathds{R}_+^2} \left(T'+\half\dda X'\cdot \ddau X'\right)+\int_{\mathds{R}}d\sigma^0 \left(\mu_B+b_2^B\ddt\dds X'\cdot \ddt\dds X' \right)\line(0,+1){9}\line(0,-1){8}_{\{\sigma^1=0\}} ~,
\end{align}
up to higher terms in $T^{-1}$, and with the corrected tension $T'=T+\Delta T_B$ and field renormalization $X'=X(1+\frac{\Delta T}{2T})$
(there are also the fermionic contributions to these, that are known \cite{Aharony:2009}, as well as to $b_2$, that we have ignored).
In the Neumann case for the $X$'s we find
\begin{align}\label{IO_2;6}
I_2&=\frac{1}{2}\int_{\mathds{R}_+^2}d^2kd^2k'\tilde X(k)\tilde X(k')k^2\left[\delta^{(2)}(k+k')+\delta^{(2)}(\bar k+k')\right]\frac{\Delta T_B}{T}\,+\nn\\
&\hspace{1.5cm}+\frac{2}{\pi}\int_{\mathds{R}_+^2}d^2kd^2k'\tilde X(k)\tilde X(k')k_0^2\delta(k_0+k'_0)\left(a_1^B-k_1^2a_2^B\right)\nn\\
&=\frac{\Delta T_B}{T}\int_{\mathds{R}_+^2} \half \dda X\cdot \ddau X+\int_{\mathds{R}}d\sigma^0 \left(a_1^B\ddt X\cdot \ddt X-a_2^B
\ddt^2 X\cdot\dds^2 X\right)
~\line(0,+1){9}\line(0,-1){10}_{\{\sigma^1=0\}}~,
\end{align}
with:
\begin{align}\label{a's result}
a_1^B=-\frac{1}{16T}\left[\sum_a m_a-\sum_{a'} m_{a'}\right]~,~a_2^B=-\frac{1}{64T}\left[\sum_a \frac{1}{m_a}-\sum_{a'} \frac{1}{m_{a'}}\right]~.
\end{align}
The resulting effective action (after the use of the free e.o.m. $\ddt^2+\dds^2=0$) is
\begin{align}\label{IO_2;7}
S_{eff}&=\int_{\mathds{R}_+^2} \left(T'+\half\dda X'\cdot \ddau X'\right)+\int_{\mathds{R}}d\sigma^0 \left(\mu_B+a_1^B\ddt X'\cdot \ddt X'+a_2^B\ddt^2 X'\cdot \ddt^2 X' \right)\line(0,+1){9}\line(0,-1){8}_{\{\sigma^1=0\}} ~.
\end{align}


\begin{thebibliography}{29}
\bibitem{Abrikosov:1956sx}
  A.~A.~Abrikosov,
  ``On the Magnetic properties of superconductors of the second group,''
  Sov.\ Phys.\ JETP {\bf 5}, 1174-1182 (1957).

\bibitem{Nielsen:1973cs}
  H.~B.~Nielsen and P.~Olesen,
  ``Vortex-line models for dual strings,''
  Nucl.\ Phys.\  B {\bf 61}, 45 (1973).


\bibitem{Arvis:1983fp}
  J.~F.~Arvis,
  ``The Exact Q Anti-Q Potential In Nambu String Theory,''
  Phys.\ Lett.\  B {\bf 127}, 106 (1983).

\bibitem{Caselle:2002rm}
  M.~Caselle, M.~Panero and P.~Provero,
  ``String effects in Polyakov loop correlators,''
  JHEP {\bf 0206}, 061 (2002)
  \href{http://arxiv.org/abs/hep-lat/0205008v1}{[arXiv:hep-lat/0205008]}.

\bibitem{Luscher:2002qv}
  M.~L\"{u}scher and P.~Weisz,
  ``Quark confinement and the bosonic string,''
  JHEP {\bf 0207}, 049 (2002)
  \href{http://arxiv.org/abs/hep-lat/0207003}{[arXiv:hep-lat/0207003]}.

\bibitem{Caselle:2002vq}
  M.~Caselle, M.~Panero, P.~Provero {\it et al.},
  ``String effects in Polyakov loop correlators,''
  Nucl.\ Phys.\ Proc.\ Suppl.\  {\bf 119}, 499-501 (2003)
  \href{http://arxiv.org/abs/hep-lat/0210023}{[arXiv:hep-lat/0210023]}.

\bibitem{Caselle:2002ah}
  M.~Caselle, M.~Hasenbusch and M.~Panero,
  ``String effects in the 3d gauge Ising model,''
  JHEP {\bf 0301}, 057 (2003)
  \href{http://arxiv.org/abs/hep-lat/0211012v2}{[arXiv:hep-lat/0211012]}.

\bibitem{Caselle:2003rq}
  M.~Caselle, M.~Hasenbusch, M.~Panero,
  ``Effective string picture for confinement at finite temperature: Theoretical predictions and high precision numerical results,''
  Nucl.\ Phys.\ Proc.\ Suppl.\  {\bf 129}, 593-595 (2004)
  \href{http://arxiv.org/abs/hep-lat/0309147}{[arXiv:hep-lat/0309147]}.

\bibitem{Caselle:2003db}
  M.~Caselle, M.~Panero, M.~Hasenbusch,
  ``Effective string picture for confining gauge theories at finite temperature,''
  \href{http://arxiv.org/abs/hep-lat/0312005}{[arXiv:hep-lat/0312005]}.

\bibitem{Caselle:2004jq}
  M.~Caselle, M.~Hasenbusch, M.~Panero,
  ``Short distance behavior of the effective string,''
  JHEP {\bf 0405}, 032 (2004)
  \href{http://arxiv.org/abs/hep-lat/0403004}{[arXiv:hep-lat/0403004]}.

\bibitem{Caselle:2004er}
  M.~Caselle, M.~Pepe, A.~Rago,
  ``Static quark potential and effective string corrections in the (2+1)-d SU(2) Yang-Mills theory,''
  JHEP {\bf 0410}, 005 (2004)
  \href{http://arxiv.org/abs/hep-lat/0406008}{[arXiv:hep-lat/0406008]}.

\bibitem{Luscher:2004ib}
  M.~L\"{u}scher and P.~Weisz,
  ``String excitation energies in SU(N) gauge theories beyond the  free-string
  approximation,''
  JHEP {\bf 0407}, 014 (2004)
  \href{http://arxiv.org/abs/hep-th/0406205}{[arXiv:hep-th/0406205]}.

\bibitem{Caselle:2005xy}
  M.~Caselle, M.~Hasenbusch, M.~Panero,
  ``Comparing the Nambu-Goto string with LGT results,''
  JHEP {\bf 0503}, 026 (2005)
  \href{http://arxiv.org/abs/hep-lat/0501027}{[arXiv:hep-lat/0501027]}.

\bibitem{Billo:2005iv}
  M.~Billo, M.~Caselle,
  ``Polyakov loop correlators from D0-brane interactions in bosonic string theory,''
  JHEP {\bf 0507}, 038 (2005)
  \href{http://arxiv.org/abs/hep-th/0505201}{[arXiv:hep-th/0505201]}.

\bibitem{Caselle:2005vq}
  M.~Caselle, M.~Hasenbusch, M.~Panero,
  ``On the effective string spectrum of the tridimensional Z(2) gauge model,''
  JHEP {\bf 0601}, 076 (2006)
  \href{http://arxiv.org/abs/hep-lat/0510107}{[arXiv:hep-lat/0510107]}.

\bibitem{Billo:2005ej}
  M.~Billo, M.~Caselle, M.~Hasenbusch {\it et al.},
  ``QCD string from D0 branes,''
  PoS {\bf LAT2005}, 309 (2006)
  \href{http://arxiv.org/abs/hep-lat/0511008}{[arXiv:hep-lat/0511008]}.

\bibitem{HariDass:2005we}
  N.~D.~Hari Dass, P.~Majumdar,
  ``High accuracy simulations of d=4 SU(3) QCD-string,''
  PoS {\bf LAT2005}, 312 (2006)
  \href{http://arxiv.org/abs/hep-lat/0511055}{[arXiv:hep-lat/0511055]}.

\bibitem{HariDass:2006pq}
  N.~D.~Hari Dass, P.~Majumdar,
  ``String-like behaviour of 4-D SU(3) Yang-Mills flux tubes,''
  JHEP {\bf 0610}, 020 (2006)
  \href{http://arxiv.org/abs/hep-lat/0608024}{[arXiv:hep-lat/0608024]}.

\bibitem{HariDass:2007tx}
  N.~D.~Hari Dass, P.~Majumdar,
  ``Continuum limit of string formation in 3-d SU(2) LGT,''
  Phys.\ Lett.\  {\bf B658}, 273-278 (2008)
\href{http://arxiv.org/abs/hep-lat/0702019}{[arXiv:hep-lat/0702019]}.

\bibitem{Brandt:2007iw}
  B.~B.~Brandt, P.~Majumdar,
  ``Luscher-Weisz algorithm for excited states of the QCD flux-tube,''
  PoS {\bf LAT2007}, 027 (2007).
\href{http://arxiv.org/PS_cache/arxiv/pdf/0709/0709.3379v1.pdf}
{[arXiv:0709.3379 [hep-lat]]}.

\bibitem{Brandt:2009tc}
  B.~B.~Brandt, P.~Majumdar,
  ``Spectrum of the QCD flux tube in 3d SU(2) lattice gauge theory,''
  Phys.\ Lett.\  {\bf B682}, 253-258 (2009).
  \href{http://arxiv.org/PS_cache/arxiv/pdf/0905/0905.4195v3.pdf}
{[arXiv:0905.4195 [hep-lat]]}.

\bibitem{Lucini:2002wg}
  B.~Lucini, M.~Teper,
  ``SU(N) gauge theories in (2+1)-dimensions: Further results,''
  Phys.\ Rev.\  {\bf D66}, 097502 (2002)
  \href{http://arxiv.org/abs/hep-lat/0206027}{[arXiv:hep-lat/0206027]}.

\bibitem{Meyer:2004hv}
  H.~Meyer, M.~Teper,
  ``Confinement and the effective string theory in SU($N \rightarrow \infty$): A Lattice study,''
  JHEP {\bf 0412}, 031 (2004)
  \href{http://arxiv.org/abs/hep-lat/0411039}{[arXiv:hep-lat/0411039]}.

\bibitem{Lottini:2005ya}
  S.~Lottini, F.~Gliozzi,
  ``The Glue-ball spectrum of pure percolation,''
  PoS {\bf LAT2005}, 292 (2006)
  \href{http://arxiv.org/abs/hep-lat/0510034}{[arXiv:hep-lat/0510034]}.

\bibitem{Caselle:2006dv}
  M.~Caselle, M.~Hasenbusch, M.~Panero,
  ``High precision Monte Carlo simulations of interfaces in the three-dimensional ising model: A Comparison with the Nambu-Goto effective string model,''
  JHEP {\bf 0603}, 084 (2006)
  \href{http://arxiv.org/abs/hep-lat/0601023}{[arXiv:hep-lat/0601023]}.

\bibitem{Billo:2006zg}
  M.~Billo, M.~Caselle, L.~Ferro,
  ``The Partition function of interfaces from the Nambu-Goto effective string theory,''
  JHEP {\bf 0602}, 070 (2006)
  \href{http://arxiv.org/abs/hep-th/0601191}{[arXiv:hep-th/0601191]}.

\bibitem{Caselle:2007yc}
  M.~Caselle, M.~Hasenbusch, M.~Panero,
  ``The Interface free energy: Comparison of accurate Monte Carlo results for the 3D Ising model with effective interface models,''
  JHEP {\bf 0709}, 117 (2007)
  \href{http://arxiv.org/abs/arXiv:0707.0055}{[arXiv:0707.0055 [hep-lat]]}.

\bibitem{Billo:2007fm}
  M.~Billo, M.~Caselle, L.~Ferro,
  ``Universal behaviour of interfaces in 2d and dimensional reduction of Nambu-Goto strings,''
  Nucl.\ Phys.\  {\bf B795}, 623-634 (2008)
  \href{http://arxiv.org/abs/arXiv:0708.3302}{[arXiv:0708.3302 [hep-th]]}.

\bibitem{Athenodorou:2007du}
  A.~Athenodorou, B.~Bringoltz, M.~Teper,
  ``The Closed string spectrum of SU(N) gauge theories in 2+1 dimensions,''
  Phys.\ Lett.\  {\bf B656}, 132-140 (2007)
  \href{http://arxiv.org/abs/arXiv:0709.0693}{[arXiv:0709.0693 [hep-lat]]}.

\bibitem{Athenodorou:2007ry}
  A.~Athenodorou, B.~Bringoltz, M.~Teper,
  ``The Spectrum of closed loops of fundamental flux in D = 2+1 SU(N) gauge theories,''
  PoS {\bf LAT2007}, 288 (2007)
  \href{http://arxiv.org/abs/arXiv:0709.2981}{[arXiv:0709.2981 [hep-lat]]}.

\bibitem{Billo:2007cw}
  M.~Billo, M.~Caselle, L.~Ferro, M.~Hasenbusch and M.~Panero,
  ``Effective string theory description of the interface free energy,''
  PoS {\bf LAT2007}, 294 (2007)
  \href{http://arxiv.org/abs/0710.1751}{[arXiv:0710.1751 [hep-lat]]}.

\bibitem{Giudice:2008zk}
  P.~Giudice, F.~Gliozzi, S.~Lottini,
  ``Confining string beyond the free approximation: The Case of random percolation,''
  PoS {\bf LATTICE2008}, 264 (2008)
  \href{http://arxiv.org/abs/arXiv:0811.2879}{[arXiv:0811.2879 [hep-lat]]}.

\bibitem{Giudice:2009di}
  P.~Giudice, F.~Gliozzi, S.~Lottini,
  ``The Confining string beyond the free-string approximation in the gauge dual of percolation,''
  JHEP {\bf 0903}, 104 (2009)
  \href{http://arxiv.org/abs/arXiv:0901.0748}{[arXiv:0901.0748 [hep-lat]]}.

\bibitem{Athenodorou:2009ms}
  A.~Athenodorou, B.~Bringoltz, M.~Teper,
  ``The Spectrum of closed loops of fundamental flux in D = 3+1 SU(N) gauge theories,''
  \href{http://arxiv.org/abs/arXiv:0912.3238}{[arXiv:0912.3238 [hep-lat]]}.

\bibitem{Athenodorou:2010cs}
  A.~Athenodorou, B.~Bringoltz, M.~Teper,
  ``Closed flux tubes and their string description in D=3+1 SU(N) gauge theories,''
  \href{http://arxiv.org/abs/arXiv:1007.4720}{[arXiv:1007.4720 [hep-lat]]}.

\bibitem{Luscher:1980fr}
  M.~L\"uscher, K.~Symanzik and P.~Weisz,
  ``Anomalies Of The Free Loop Wave Equation In The Wkb Approximation,''
  Nucl.\ Phys.\  B {\bf 173}, 365 (1980).

\bibitem{Luscher:1980ac}
  M.~L\"{u}scher,
  ``Symmetry Breaking Aspects Of The roughening Transition In Gauge Theories,''
  Nucl.\ Phys.\  B {\bf 180}, 317 (1981).

\bibitem{Aharony:2009}
  O.~Aharony and E.~Karzbrun,
  ''On the effective action of confining strings,''
  JHEP {\bf 0906}, 012 (2009)
  \href{http://arxiv.org/abs/0903.1927}{[arXiv:0903.1927 [hep-th]]}.

\bibitem{Meyer:2006qx}
  H.~B.~Meyer,
  ``Poincare invariance in effective string theories,''
  JHEP {\bf 0605}, 066 (2006)
  \href{http://arxiv.org/abs/hep-th/0602281}{[arXiv:hep-th/0602281]}.

\bibitem{Aharony:2010}
  O.~Aharony, Z.~Komargodski and A.~Schwimmer,
  work in progress, presented by O. Aharony at the Strings 2009 conference, June 2009,
  {\tt http://strings2009.roma2.infn.it/talks/Aharony\_Strings09.ppt}, and
  at the ECT* workshop on ``Confining flux tubes and strings'', July 2010,
  {\tt http://www.ect.it/Meetings/ConfsWksAndCollMeetings/ConfWksDocument/2010/ talks/Workshop\_05\_07\_2010/Aharony.ppt}.

\bibitem{Braaten:1986bz}
  E.~Braaten, R.~D.~Pisarski and S.~M.~Tse,
  ``The Static potential for smooth strings,''
  Phys.\ Rev.\ Lett.\  {\bf 58}, 93 (1987)
  [Erratum-ibid.\  {\bf 59}, 1870 (1987)].

\bibitem{Braaten:1987gq}
  E.~Braaten and S.~M.~Tse,
  ``The Static potential for smooth strings in the large D limit,''
  Phys.\ Rev.\  D {\bf 36}, 3102 (1987).

\bibitem{Polyakov:1986cs}
  A.~M.~Polyakov,
  ``Fine Structure of Strings,''
  Nucl.\ Phys.\  B {\bf 268}, 406 (1986).

\bibitem{Maldacena:1997re}
  J.~M.~Maldacena,
  ``The large N limit of superconformal field theories and supergravity,''
  Adv.\ Theor.\ Math.\ Phys.\  {\bf 2}, 231 (1998)
  [Int.\ J.\ Theor.\ Phys.\  {\bf 38}, 1113 (1999)]
  \href{http://arxiv.org/abs/hep-th/9711200}{[arXiv:hep-th/9711200]}.

\bibitem{Aharony:2010a}
  O.~Aharony and N. Klinghoffer, ``Corrections to Nambu-Goto energy levels from
  the effective string action,'' to appear.

\bibitem{Dietz:1982uc}
  K.~Dietz and T.~Filk,
  ``On The renormalization Of String Functionals,''
  Phys.\ Rev.\  D {\bf 27}, 2944 (1983).

\bibitem{Nesterenko:1997ku}
  V.~V.~Nesterenko and I.~G.~Pirozhenko,
  ``Justification of the zeta function renormalization in rigid string
  model,''
  J.\ Math.\ Phys.\  {\bf 38}, 6265 (1997)
  \href{http://arxiv.org/abs/hep-th/9703097}{[arXiv:hep-th/9703097]}.

\bibitem{Caselle:1996kd}
  M.~Caselle and K.~Pinn,
  ``On the Universality of Certain Non-Renormalizable Contributions in
  Two-Dimensional Quantum Field Theory,''
  Phys.\ Rev.\  D {\bf 54}, 5179 (1996)
  \href{http://arxiv.org/abs/hep-lat/9602026}{[arXiv:hep-lat/9602026]}.

\bibitem{Peskin:1995ev}
  M.~E.~Peskin and D.~V.~Schroeder,
  ``An Introduction To Quantum Field Theory,''
\href{http://www.slac.stanford.edu/spires/find/hep/www?irn=3485960}
{\it  Reading, USA: Addison-Wesley (1995)}

\bibitem{Witten:1998zw}
  E.~Witten,
  ``Anti-de Sitter space, thermal phase transition, and confinement in  gauge
  theories,''
  Adv.\ Theor.\ Math.\ Phys.\  {\bf 2}, 505 (1998)
  \href{http://arxiv.org/abs/hep-th/9803131}{[arXiv:hep-th/9803131]}.

\bibitem{Maldacena:2000yy}
  J.~M.~Maldacena and C.~Nunez,
  ``Towards the large N limit of pure N = 1 super Yang Mills,''
  Phys.\ Rev.\ Lett.\  {\bf 86}, 588 (2001)
  \href{http://arxiv.org/abs/hep-th/0008001}{[arXiv:hep-th/0008001]}.

\bibitem{Klebanov:2000hb}
  I.~R.~Klebanov and M.~J.~Strassler,
  ``Supergravity and a confining gauge theory: Duality cascades and
  chiSB-resolution of naked singularities,''
  JHEP {\bf 0008}, 052 (2000)
 \href{http://arxiv.org/abs/hep-th/0007191}{[arXiv:hep-th/0007191]}.

\bibitem{Bertoldi:2004rn}
  G.~Bertoldi, F.~Bigazzi, A.~L.~Cotrone, C.~Nunez and L.~A.~Pando Zayas,
  ``On the universality class of certain string theory hadrons,''
  Nucl.\ Phys.\  B {\bf 700}, 89 (2004)
  \href{http://arxiv.org/abs/hep-th/0401031}{[arXiv:hep-th/0401031]}.

\end{thebibliography}
\end{document}